\documentclass[amsmath, amssymb, showkeys, titlepage, superscriptaddress, preprint]{revtex4-2}

\usepackage{mathpazo}
\usepackage{booktabs}
\usepackage{graphicx}
\usepackage{rotating}
\usepackage{multirow}
\usepackage{xr-hyper}
\usepackage[bookmarks=true, colorlinks=true, citecolor=black, linkcolor=black, urlcolor=black]{hyperref}
\usepackage{xcolor}
\usepackage{tabularray}
\usepackage{siunitx}
\usepackage{setspace}
\usepackage{orcidlink}

\begin{document}

\title{Mapping the changing structure of science through diachronic periodical embeddings}

\author{Zhuoqi Lyu\orcidlink{0000-0002-5787-0862}}
\affiliation{Department of Data Science, College of Computing, City University of Hong Kong, Hong Kong, China}

\author{Qing Ke\orcidlink{0000-0002-2945-5274}}
\email{Correspondence to q.ke@cityu.edu.hk}
\affiliation{Department of Data Science, College of Computing, City University of Hong Kong, Hong Kong, China}

\date{\today}

\begin{abstract}
Understanding the changing structure of science over time is essential to elucidating how science evolves. We develop diachronic embeddings of scholarly periodicals to quantify ``semantic changes'' of periodicals across decades, allowing us to track the evolution of research topics and identify rapidly developing fields. By mapping periodicals within a physical-life-health triangle, we reveal an evolving interdisciplinary science landscape, finding an overall trend toward specialization for most periodicals but increasing interdisciplinarity for bioscience periodicals. Analyzing a periodical's trajectory within this triangle over time allows us to visualize how its research focus shifts. Furthermore, by monitoring the formation of local clusters of periodicals, we can identify emerging research topics such as AIDS research and nanotechnology in the 1980s. Our work offers novel quantification in the science of science and provides a quantitative lens to examine the evolution of science, which may facilitate future investigations into the emergence and development of research fields. 
\end{abstract}

\keywords{science of science $|$ evolution of science $|$ map of science $|$ word2vec}

\maketitle

\section{Introduction}

Since the establishment of the first scholarly journal, \emph{Philosophical Transactions of the Royal Society}, in 1665~\cite{henry1665epistle}, journals and conference proceedings have served as the primary outlet for science publishing, crucial for the dissemination of research findings within the scientific community~\cite{baldwin2015making, csiszar2018scientific}. These periodicals are also instrumental in shaping scientific norms; publishing in them signals affiliations with and establishes standings within a particular scientific community. Additionally, they play a major role in announcing the priority of scientific discoveries, which may influence eligibility for awards and recognition.  

As scholarly periodicals tend to publish thematically coherent sets of papers, they are widely considered as a viable representation of knowledge components and consequently used in numerous inquiries into the scientific enterprise. These include identifying recombinant innovation in science~\cite{uzzi2013atypical, wang2017bias}, categorizing biomedical research~\cite{narin1976structure}, and assessing the interdisciplinary integration of emerging fields~\cite{nunez2019happened}, among others~\cite{line1970half, calcagno2012flows}. Notably, periodicals have long been used as instruments to probe the structure of science~\cite{leydesdorff1987various, small1999visualizing, shiffrin2004mapping}, by representing them as crisp discipline vectors, sparse vectors that capture (co-)citation relationships~\cite{narin1976structure, boyack2005mapping, rosvall2008maps, borner2012design} or online activities~\cite{bollen2009clickstream}, as well as dense vectors based on representation learning methods~\cite{peng2021neural}. These extensive efforts have yielded several global maps of science that are useful for understanding knowledge flow between fields and supporting decision-making processes such as portfolio analysis and resource allocation. 

However, these maps are static and fail to capture the dynamic nature of both the evolution of science and the development of scholarly communication. Specifically, science is continually evolving, with new fields and research topics emerging sporadically, which can reshape the landscape of science publishing. For example, the recognition of AIDS as a new disease in the 1980s led to the creation of several journals to accommodate the unmet need for platforms for publishing AIDS-related research. In addition, the rise of interdisciplinary science in recent decades has likely brought certain fields closer to each other~\cite{gates2019nature}. In terms of scholarly communication, technological advancements have drastically transformed science publishing: It is now common for researchers to read and search scientific literature online; many periodicals have transitioned from the printed to the online medium, which largely eliminates space constraints and allows for more extensive referencing and publishing. All these developments cannot be captured by static, retrospective vectors of periodicals, highlighting the need for dynamic, time-varying representations. 

Here, we develop diachronic embeddings of periodicals to examine how these embeddings evolve over time, building on our previous methodology~\cite{peng2021neural}. Previous studies related to ours have studied relationships between disciplines over time using vector representations of disciplines~\cite{mcgillivray2022investigating}, without focusing on the more fine-grained level of periodicals. We demonstrate that, by using diachronic embeddings, we can quantify the magnitude of ``semantic change'' for each periodical based on its nearest neighbors at different times, chart the direction of semantic changes, and track the evolving frontier of interdisciplinary periodicals. Moreover, by tracking the formation of local clusters of periodicals, we identify numerous emerging themes, such as AIDS research and biomaterials. Overall, our work sheds light on the changing structure of science at the periodical level and provides insights into the evolution of science.

\section{Results}

\subsection{Constructing diachronic periodical embeddings}

To build diachronic embeddings of periodicals (see Fig.~S2 for a schematic illustration), we begin by constructing a citation network for each decade $t$ starting from the 1950s. In the network, nodes represent papers published during that decade and directed links point from citing papers to cited ones. For each network, we perform random walks on it to generate paper citation trails and then map each paper in a citation trail to the periodical in which it was published, resulting in periodical level trails (see Methods). By generating a large number of citation trails, we create an effective exploration of the citation network. Treating each trail as a ``sentence'' and each periodical as a ``word'', we apply word2vec~\cite{mikolov2013efficient} to the periodical trail corpus to learn vector representations of periodicals, denoted as $\mathbf{v}_i^t$ for periodical $i$. Similar to word embeddings, periodicals with similar ``contextual'' periodicals in citation trails are closer in the vector space, reflecting their semantic similarity. For example, in the 2010s, the periodicals most similar to \emph{Nature} are \emph{Science}, \emph{Nature Communication}, and \emph{Science Advances}, highlighting their multidisciplinarity feature. However, back in the 1950s, \emph{Nature} was largely a biology journal and was most similar to \emph{Journal of Molecular Biology}, \emph{Biochimica et Biophysica Acta}, and \emph{Naturwissenschaften} (see Table~S5 for the top neighbors of \emph{Nature} over time). 

\begin{figure*}[t!]
\centering
\includegraphics[trim=0 5mm 0 0, width=.9\linewidth]{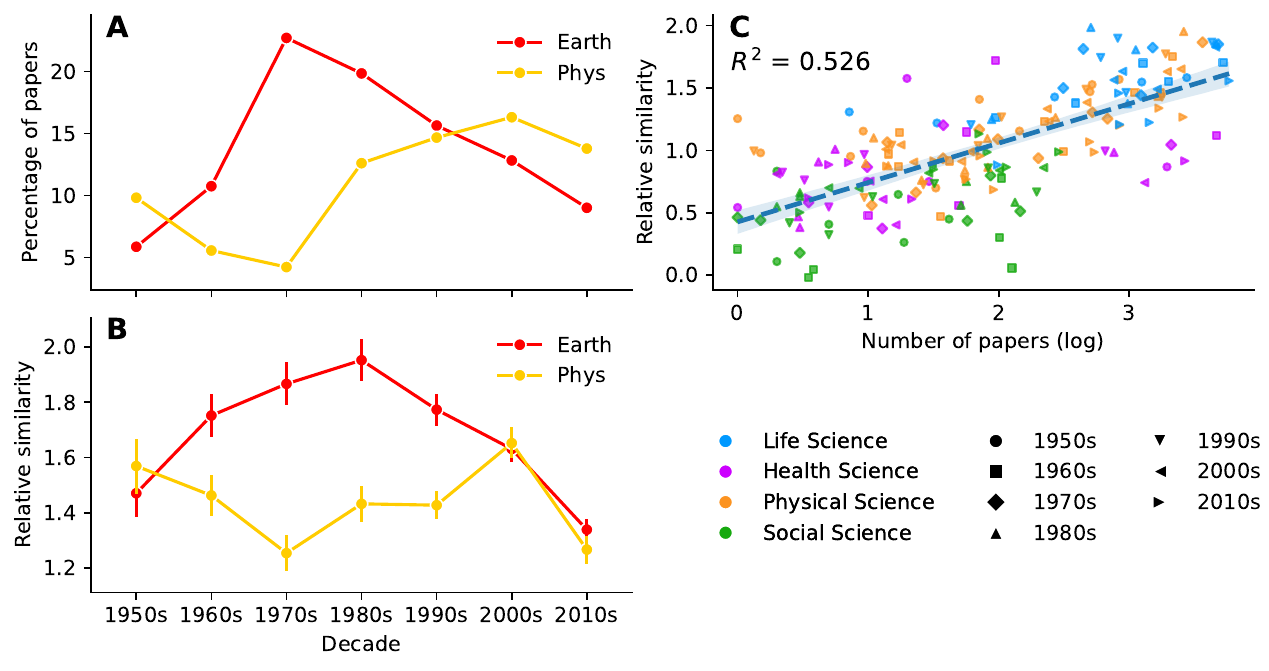}
\caption{\textbf{Validating diachronic embeddings using \emph{Nature}.} 
(A) Percentage of papers in Earth and Planetary Sciences and Physics published in \emph{Nature} by decade. Papers in the 2010s refer to those published in 2010--2021 for simplicity. 
(B) Relative similarity between \emph{Nature} and the two focused disciplines. Relative similarity is defined as the average cosine similarity between \emph{Nature} and all periodicals belonging to that discipline, divided by the average cosine similarity between \emph{Nature} and all periodicals. 
(C) The correspondence between publication volume and relative similarity. Color represents discipline and shape marks decade.}
\label{fig:validation}
\end{figure*}

We validate our diachronic periodical embeddings using case periodicals. First, we focus on multidisciplinary journals, as they publish papers in multiple disciplines and disciplinary compositions of published papers over time should correlate with their similarities to disciplines, given that citation exchanges tend to occur within disciplines. For example, by re-assigning each \emph{Nature} paper to the field corresponding to the majority of its references' fields, we observe a rapid increase in the publication volume of papers in Earth and Planetary Sciences (a Scopus field label) during the 1970s--1990s (Fig.~\ref{fig:validation}A). This makes \emph{Nature} more likely to occupy similar positions in citation trails as geoscience periodicals and thus more similar to the field. Indeed, when we measure the closeness between \emph{Nature} and the geoscience field by calculating the average cosine similarity to all periodicals in the field relative to the average across all periodicals, we find that \emph{Nature} becomes closer to geoscience during the same period (Fig.~\ref{fig:validation}B). In general, there is a positive correlation between publication volume in a field and relative cosine similarity (coefficient of determination $R^2 = 0.526$; Fig.~\ref{fig:validation}C), suggesting that our diachronic embeddings can be used to quantify how the ``semantics'' of \emph{Nature} change over time. Repeating this analysis for two other multidisciplinary journals, \emph{Science} and \emph{PNAS}, we find a similar correspondence (Figs.~S10--S11), further supporting the validity of our diachronic periodical embeddings. 

Second, in a similar vein, we examine \emph{Cognitive Science}, a flagship journal of the field of cognitive science. This field was established in the 1950s with the vision that fruitful cross-fertilization among six diverse disciplines---psychology, linguistics, artificial intelligence, anthropology, philosophy, and neuroscience---would advance the science of mind~\cite{nunez2019happened}. However, both commentaries and empirical analyses have noted that over the course of its development, cognitive science has lost its intended diversity and instead been characterized by an overrepresentation of psychology. Our analysis using diachronic embeddings supports this observation; the proportion of psychology papers published in \emph{Cognitive Science} has increased significantly, and the journal's closeness to psychology has also increased (Fig.~S12). 

\subsection{Quantifying semantic change of periodicals}

\begin{figure*}[t!]
\centering
\includegraphics[trim=0 8mm 0 0, width=.9\linewidth]{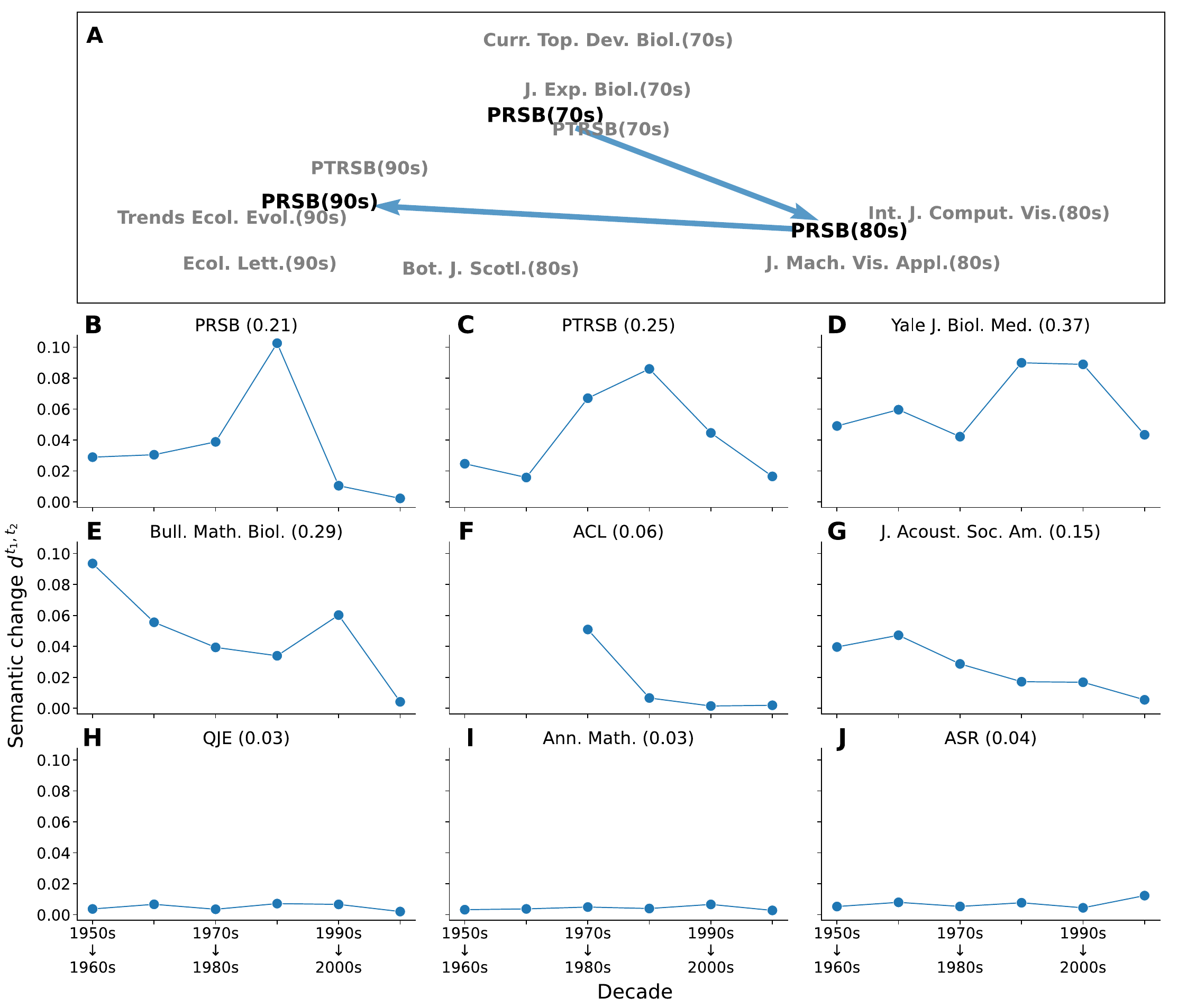}
\caption{\textbf{Quantifying semantic change, $d^{t_1,t_2}$, of a periodical.} 
(A) Two-dimensional visualization of \emph{PRSB}'s semantic change based on its diachronic embeddings. During the 1970s--1990s, it shifted from a cluster of biology periodicals to computer vision to ecology. 
(B--J) $d^{t_1,t_2}$ for individual periodicals over time. Numbers in parentheses in the titles are total $d^{t_1,t_2}$ over time. Figs.~S14--S15 provide more examples.}
\label{fig:SC}
\end{figure*}

Our validation exercises above indicate that there are semantic changes for certain periodicals, raising the questions of how to quantify the magnitude of these changes and which periodicals have undergone the most significant transformations. Inspired by computational linguistics studies on semantic changes of words~\cite{shoemark2019room}, we quantify a periodical's semantic changes over time by looking at its $k$ nearest neighbors based on its diachronic embeddings (Fig.~\ref{fig:SC}A). Specifically, let $N_i^{t_1, t_2} = N_i^{t_1} \cup N_i^{t_2}$ represent the union of periodical $i$'s $k$-nn at $t_1$ and $t_2$ (with $k$ set to 10). We create a vector $s^{t_1}$ for $t_1$, where each entry is the cosine similarity between $\mathbf{v}_i^{t_1}$ and $\mathbf{v}_n^{t_1}$ ($n \in N_i^{t_1, t_2}$), and similarly create another vector $s^{t_2}$ for $t_2$. We then measure the semantic change of $i$ from $t_1$ to $t_2$, denoted as $d_i^{t_1,t_2}$, as the cosine distance between $s^{t_1}$ and $s^{t_2}$, \emph{i.e.}, $d_i^{t_1,t_2} = 1 - \frac{s^{t_1} \cdot s^{t_2}}{||s^{t_1}|| \cdot ||s^{t_2}||}$. A periodical tends to have a large $d$ if there is a low overlap between $N_i^{t_1}$ and $N_i^{t_2}$. 

By calculating $d$ between two consecutive decades, we find that across time, periodicals tend to experience limited semantic changes, with the median and the 95th percentile of $d$ around 0.01 and 0.04, respectively (Fig.~S13). For example, the semantics of \emph{Quarterly Journal of Economics}, \emph{Annals of Mathematics}, and \emph{American Sociological Review} have remained almost unchanged over the past seven decades (Figs.~\ref{fig:SC}H--J). On the other hand, a few periodicals have undergone drastic semantic shifts. Notably, the \emph{Proceedings of the Royal Society B: Biological Science} (\emph{PRSB}) experienced a semantic drift in the 1980s, during which it moved closer to computer vision journals like \emph{International Journal of Computer Vision} and \emph{Journal of Machine Vision and Applications} (Fig.~\ref{fig:SC}A). This was due to the publication of a few highly influential computer vision papers~\cite{marr1980edgedetection, LonguetHiggins1980movingretinal, Srinivasan1982predictivecoding, Perrett1985visualcells, Morrone1988featuredetection, Marr1981Directionalselectivity}, which attracted volumes of citations from this field. Subsequently, \emph{PRSB} returned to a focus on biology, and its semantic changes decreased over the following decades (Fig.~\ref{fig:SC}B). Similarly, its sister journal, \emph{Philosophical Transactions of the Royal Society B}, experienced a comparable semantic journey (Fig.~\ref{fig:SC}C), largely due to studies published in the 1980s at the intersection of neural systems and computations~\cite{gregory1980perceptions, posner1982neural, lee1980optic}, which generated numerous citations from topics outside biology, particularly in computer vision, neural networks, and cognitive science. Likewise, \emph{Yale Journal of Biology and Medicine} exhibited significant semantic changes during the 1980s--2000s (Fig.~\ref{fig:SC}D). Upon examining its neighbors over time, we find that it had a rapid shift in research interests from general medicine to digestive diseases in the 1990s, which was later restored in the 2000s. Finally, Figs.~\ref{fig:SC}E--G highlight three periodicals---\emph{Bulletin of Mathematical Biology}, \emph{Annual Meeting of the Association for Computational Linguistics}, and \emph{Journal of the Acoustical Society of America}---whose semantic trajectories have steadily decreased, indicating a stabilization of their neighbors over time. 

We further identify periodicals that have undergone the largest topical shifts by summing their semantic changes $d^{t_1,t_2}$ over time. Fig.~S16 presents the distributions of this total change for periodicals grouped by the decade of their establishment, with several periodicals highlighted within these distributions. We observe that for periodicals established before the 1960s, the distribution of total semantic changes is much flatter compared to those established in later decades. Moreover, there is a graduate emergence of a mode value for total semantic changes, suggesting a potential ubiquity of semantic change over time. 

Examining semantic changes at the field level, Fig.~S17 shows the distributions of total semantic changes for periodicals across 27 fields. The top three fields experiencing the most shifts are Multidisciplinary, Chemistry, and Biochemistry. Multidisciplinary periodicals are designed to publish contributions spanning various disciplines. Chemistry, often referred to as ``the central science'', plays a pivotal role in linking upstream physical sciences with downstream fields like life sciences and medicine~\cite{balaban2006chemistry, szell2018nobel, peng2021neural}. This unique position at the intersection of multiple scientific domains underscores its greater potential for semantic displacement. Biochemistry is also recognized as a highly interdisciplinary field~\cite{boyack2005mapping}. In contrast, specialized fields such as management, accounting, dentistry, and health professions tend to be more stable. 

\subsection{Identifying direction of semantic change}

Beyond quantifying the magnitude of semantic changes, we are also interested in understanding the direction of semantic displacement. In doing this, we define several disciplinary poles in the vector space and measure a periodical's closeness to each pole. Specifically, we focus on the three broad research areas designated by Scopus: physical science, life science, and health science, and identify the pole for each area by averaging the vectors of all periodicals in that area. Formally, let $\mathcal{P}^t$ represent the set of vectors for physical science periodicals in decade $t$. The physical science pole is the centroid of this set: $\overline{\mathbf{v}}_{\mathcal{P}}^t = \frac{1}{\left\vert \mathcal{P}^t \right\vert} \sum_{\mathbf{v}^t \in \mathcal{P}^t} \mathbf{v}^t$. We measure the closeness between a periodical $i$ and the pole as the cosine similarity between $\mathbf{v}_i^t$ and $\overline{\mathbf{v}}_{\mathcal{P}}^t$, denoted as $l_{i,\mathcal{P}}^t$. Similarly, we identify the life and health science poles, $\overline{\mathbf{v}}_{\mathcal{L}}^t$ and $\overline{\mathbf{v}}_{\mathcal{H}}^t$, and calculate periodical $i$'s closeness to the two areas, $l_{i,\mathcal{L}}^t$ and $l_{i,\mathcal{H}}^t$. We then normalize the proximity to the three areas, $(l_{i,\mathcal{P}}^t, l_{i,\mathcal{L}}^t, l_{i,\mathcal{H}}^t)$, to form a probability distribution, allowing us to place all periodicals within the physical-life-health triangle. 

\begin{figure*}[t!]
\centering
\includegraphics[trim=0 5mm 0 0, width=\linewidth]{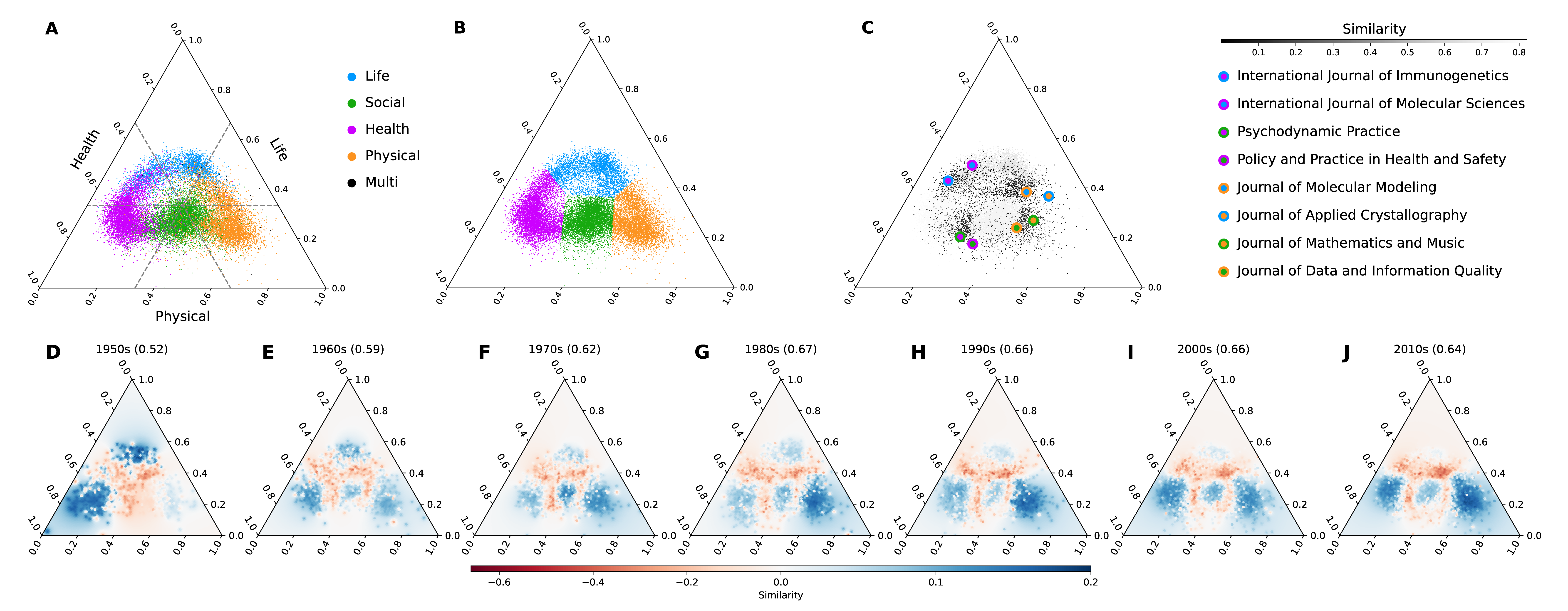}
\caption{\textbf{Mapping periodicals within the physical-life-health triangle.} 
(A) A ternary plot showing the distribution of all periodicals with respect to three conceptional axes: physical science, life Science, and health science, in the 2010s. Color denotes research area assigned by Scopus. 
(B) The same ternary plot but with periodicals colored by cluster labels generated by $k$-means based on periodicals' ternary coordinates. The label of a cluster is the most common Scopus area label. 
(C) The same ternary plot but with periodicals colored by the level of disagreement between $k$-means clustering and Scopus labels. Periodicals with larger disagreement are colored darker. Highlighted are 8 misclassified periodicals, whose central colors indicate their clustering labels and edge colors represent research areas assigned by Scopus. 
(D-J) Interpolated heatmaps of disagreement between $k$-means clustering results and Scopus labels for each decade. The interpolation is based on inverse distance weighting (IDW). Numbers in titles are average similarity of all periodicals.} 
\label{fig:ternary_scatter}
\end{figure*}

Fig.~\ref{fig:ternary_scatter}A presents the positions of all periodicals within the triangle for the 2010s, with colors representing their Scopus labels, to facilitate comparisons with this traditional journal classification system (see Fig.~S22 for other decades). In this ternary plot, a point's position indicates its relative proximity to the three research areas. For example, a periodical closer to the health science (left) corner can be characterized as a health science journal. By construction, periodicals belonging to one area are located nearer to the corresponding corner. Although social science periodicals are not used in forming the triangle, they can still be positioned within it, and they tend to be closer to the physical and health sciences than to the life science. 

Fig.~\ref{fig:ternary_scatter}A reveals blurred and indistinct regions within the triangle where periodicals with different area labels intermingle, and for some periodicals, there is a discrepancy between the manually curated Scopus discipline labels and our data-driven discipline labels based on embeddings. This exposes the nuanced structure in the disciplinary organization and underscores the limitations of categorical classification approaches, emphasizing the need to uncover interdisciplinary periodicals. To address this, we apply $k$-means clustering to group periodicals into four clusters based on their ternary coordinates (Fig.~\ref{fig:ternary_scatter}B) and compare these clusters with the four broad areas in the Scopus classification system using an element-centric measurement, which quantifies similarity between two clusterings at the individual element level~\cite{gates2019element}. Fig.~\ref{fig:ternary_scatter}C displays the same map with periodicals colored by similarity, where periodicals with lower similarity are shown in darker shades to highlight disagreements between the two clusterings. We observe that disagreements are indeed located at the intersections of disciplines. Through manual inspection, we find that misclassified journals often exhibit a higher level of interdisciplinarity, including \emph{International Journal of Immunogenetics}, \emph{Journal of Molecular Modeling}, and \emph{Journal of Mathematics and Music}. To pinpoint regions with the highest interdisciplinarity, we use an inverse distance weighting approach~\cite{shepard1968two} to generate an interpolated heatmap of similarity, shown in Fig.~\ref{fig:ternary_scatter}J. This heatmap reveals that the region closer to the life science corner exhibits a much higher level of interdisciplinarity than those closer to the physical and health science corners. 

To further characterize the shift in interdisciplinarity over time, we generate heatmaps for other decades (Figs.~\ref{fig:ternary_scatter}D--I). We observe that overall the level of disagreement has decreased over time, indicating a trend towards disciplinary cohesion. For the physical, health, and social science fields, disciplinary organization has become more pronounced, whereas life science periodicals are increasingly blending with those from other disciplines, calling for the need for adopting data-driven discipline classification systems. Correspondingly, interdisciplinary hotspots have shifted from the center of the triangle to the intersections between life science and both physical and health sciences. 

\begin{figure*}[t!]
\centering
\includegraphics[trim=0 5mm 0 0, width=0.8\linewidth]{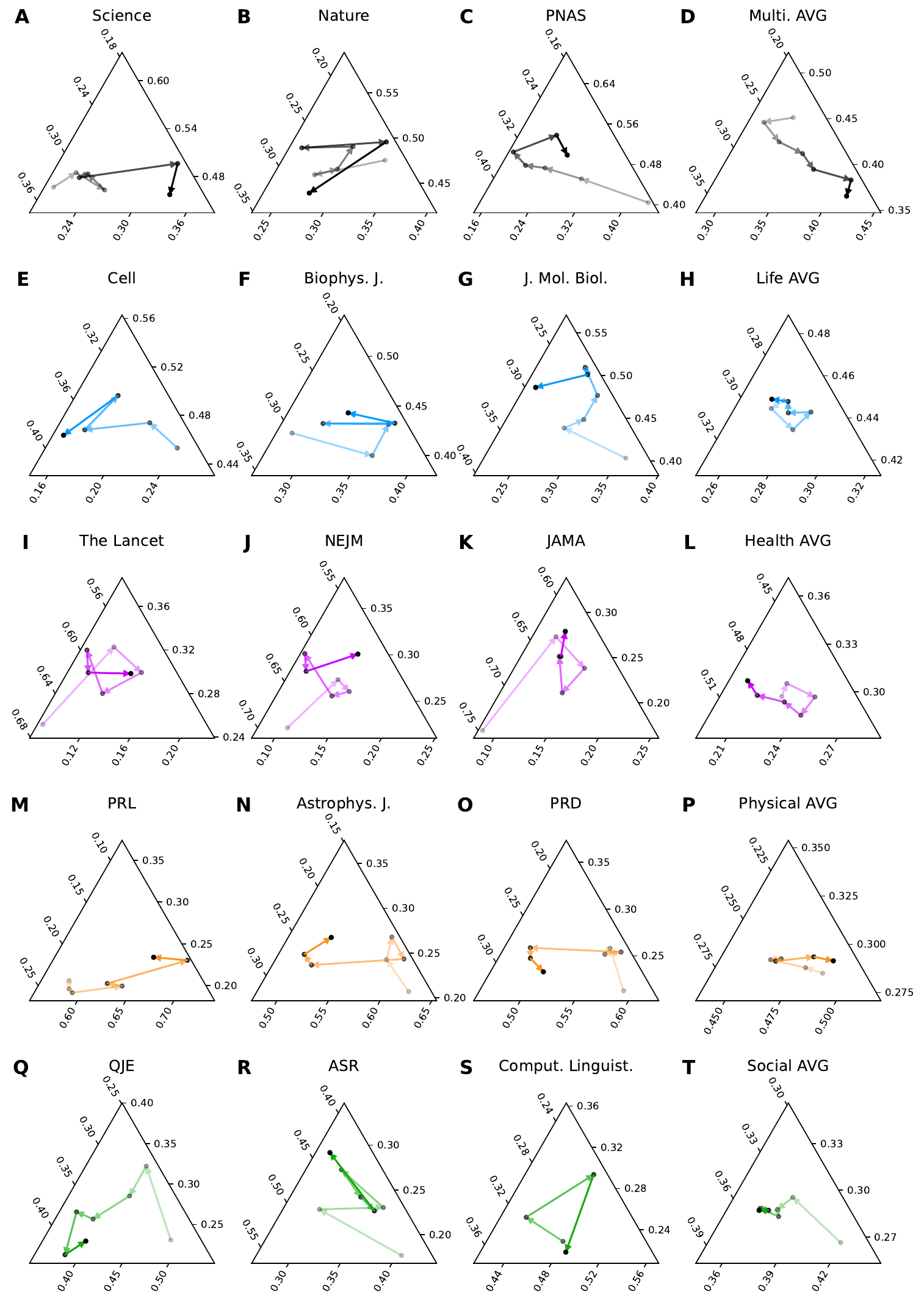}
\caption{\textbf{Charting evolution traces of periodicals within the physical-life-health triangle.} 
We show trajectories of closeness to the three research areas for 15 periodicals and the averaged trajectories over all periodicals in each category (the last column). Each trajectory is formed by sequentially connecting the positions in the triangle with arrows, from the 1950s (or the decade of establishment) to the 2010s.}
\label{fig:ternary}
\end{figure*}

Furthermore, by tracking the positions of a periodical within the triangle over time, we can obtain a trajectory that reflects how its research topics evolve temporally. Fig.~\ref{fig:ternary} presents the trajectories of 15 periodicals. Notably, \emph{Science} exhibits significant semantic displacement toward physical science particularly during the 1990s and 2000s (Fig.~\ref{fig:ternary}A), which is consistent with a substantial increase in the number of papers published in Physics and Astronomy. In contrast, \emph{Nature} has largely maintained its position over the past 70 years, although it has gradually moved toward the center of the triangle (Fig.~\ref{fig:ternary}B). This aligns with previous research indicating that \emph{Nature} has garnered citations from an increasingly diverse range of disciplines~\cite{gates2019nature}. Meanwhile, \emph{PNAS} experienced a rapid shift away from physical science, gravitating toward life science in the 1950s--1980s (Fig.~\ref{fig:ternary}C). 

Life science periodicals exhibit a variety of evolutionary trajectories. Over the past 40 years, \emph{Cell} has shifted away from physical science and gradually moved closer to health science (Fig.~\ref{fig:ternary}E). Meanwhile, \emph{Biophysical Journal} has experienced fluctuations along the physical science axis, ultimately drawing nearer to physical science (Fig.~\ref{fig:ternary}F), which reflects the increasing reliance of biology and medicine on physical science~\cite{hill1956biophysics}. The \emph{Journal of Molecular Biology} showed a great shift toward life science until the 2000s (Fig.~\ref{fig:ternary}G), which seems to be in accordance with the view that the molecular paradigm ceased to be a reliable guide for biology as it was throughout the 20th century~\cite{woese2004new}. 

Turning to health science, \emph{Lancet}, \emph{New England Journal of Medicine}, and \emph{JAMA} share a similar path of gradually becoming less health science but more life science (Figs.~\ref{fig:ternary}I--K). This trend diverges from the overall trajectory of health science periodicals and may suggest a transformative role played by these prestige periodicals in the convergence of biology and medicine, driven by an increasing dependence of medicine on upstream scientific discoveries from life science, as well as advancements in technologies and instruments for diagnosis and therapeutics~\cite{lowy2011historiography}. 

Looking at physical science periodicals, \emph{The Astrophysical Journal}, \emph{Physical Review D}, and \emph{PRL} underwent significant changes in the 1980s--1990s. The first two shifted away from physical science, whereas \emph{PRL} moved in the opposite direction (Fig.~\ref{fig:ternary}M--O). Still, they all gravitated towards life science, likely reflecting the thriving of biophysics as a distinct discipline. Finally, a noticeable trend towards life sciences was observed for the three social science periodicals (Figs.~\ref{fig:ternary}Q--S). However, \emph{QJE} and \emph{Computational Linguistics} reverted to their original levels as when they were established, while \emph{ASR} ended up closer to life science. The relatively stable trajectory of \emph{Computational Linguistics} over time may suggest its ongoing commitment to bridge computer science and linguistics. 

\subsection{Detecting emerging research topics}

As periodicals typically publish topically coherent papers, the birth of new periodicals may signal the emergence of new fields or research directions. For instance, the identification of AIDS as a novel disease in the 1980s led to the establishment of a number of new journals, including \emph{AIDS} and \emph{The Lancet HIV}. Similarly, the invention of the scanning tunneling microscope and the discovery of fullerenes during the 1980s---both of which earned Nobel Prizes in 1986 and 1996, respectively---catalyzed the growth of nanotechnology research. This resulted in the creation of numerous journals, such as \emph{Nanotechnology}, \emph{Nano Letters}, and \emph{Advanced Materials}, dedicated to this exciting new area. These observations prompt us to ask: Can we identify emerging research topics in the evolution of science through the lens of newly established periodicals? 

\begin{figure*}[t!]
\centering
\includegraphics[trim=0 2mm 0 0, width=\linewidth]{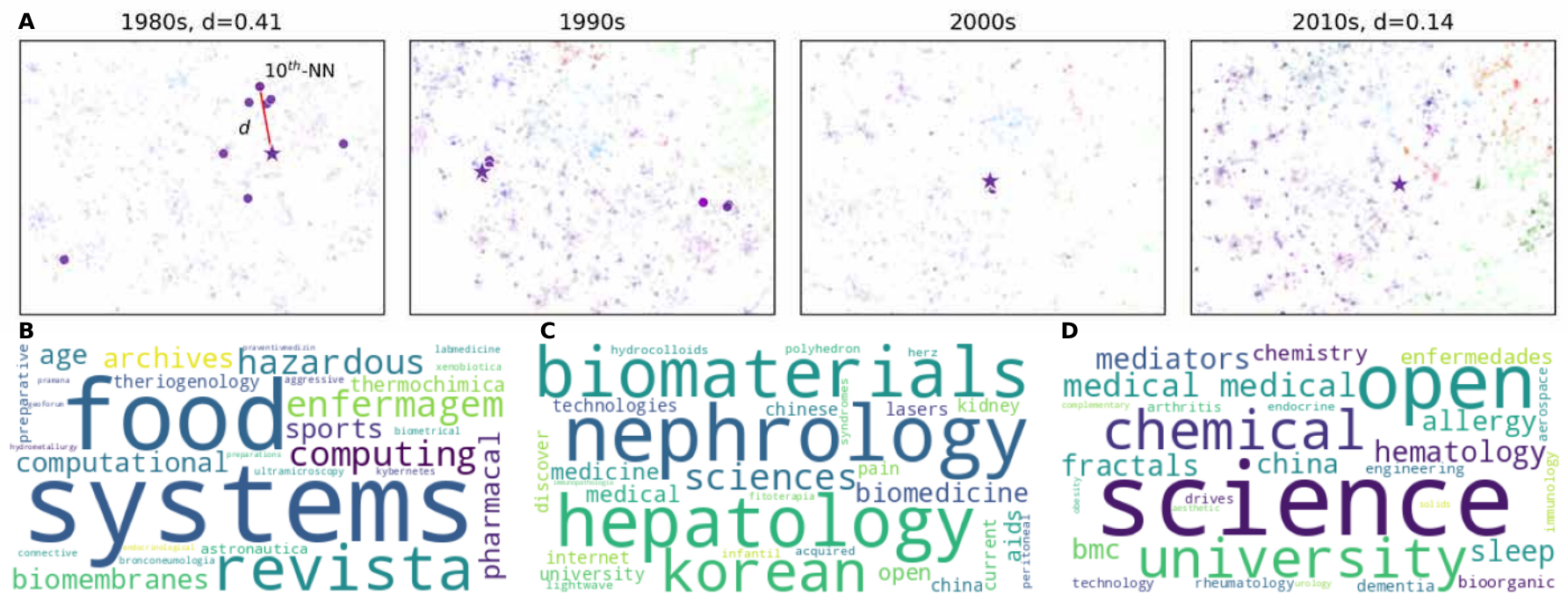}
\caption{\textbf{Detecting emerging research topics.} 
(A) 2-d visualizations of \emph{AIDS} (marked as stars) and its 10-nearest neighbors (marked as circles) in each decade. The red line marks $d$, the cosine distance from \emph{AIDS} to its 10th nearest neighbor. 
(B--D) The most representative words appeared in the titles of top 10\% periodicals based on $\Delta d$ for periodicals established in the (B) 1970s, (C) 1980s, and (D) 1990s.}
\label{fig:emergence}
\end{figure*}

Let us begin by examining individual periodicals. We hypothesize that as an emerging research topic matures, a tight cluster forms in which periodicals are highly similar to each other. For example, focusing on the journal \emph{AIDS}, Fig.~\ref{fig:emergence}A illustrates this process. It shows that the nearest neighbors of \emph{AIDS} became stable starting in the 2000s, with distances between them decreasing over time. This indicates a densification of its most similar neighbors and the formation of a locally cohesive cluster related to AIDS/HIV research, as shown by the periodical names (Table~S7). We quantify this process by calculating the change in distance to its $k$-th nearest neighbor from the decade of establishment to the last decade: $\Delta d = d^{t_1} - d^{\text{2010s}}$. For \emph{AIDS}, $\Delta d = 0.27$. A large $\Delta d$ signifies a substantial reduction in the minimum distance to cover its $k$ nearest neighbors, thereby pointing to the formation of a local cluster. 

We calculate $\Delta d$ for each new periodical across decades, allowing us to identify other periodicals that have undergone a similar process of forming a tight cluster since their inception. In the 1980s, notable examples include \emph{Sleep} ($\Delta d = 0.31$), \emph{Biomaterials} ($\Delta d = 0.29$), \emph{Journal of Controlled Release} ($\Delta d = 0.26$), and \emph{Applied Organometallic Chemistry} ($\Delta d = 0.26$) (Table~S9). Overall, the distribution of $\Delta d$ is symmetrically centered around zero, indicating that only a minority of periodicals have wither merged into clusters or dissociated themselves from their neighborhood over time (Fig.~S24). To identify potential emerging research topics, we calculate the representativeness of words in periodical titles by comparing periodicals with $\Delta d$ in the top 10\% with non-top ones~\cite{monroe2017fightin}, hypothesizing that important topics may appear in multiple periodicals with large $\Delta d$. Figs.~\ref{fig:emergence}B--D showcase the most representative words for periodicals established from the 1970s to the 1990s. In the 1970s, systems research and food science emerged as notable topics; the 1980s saw a rising interest in nephrology, hepatology, biomaterials, and AIDS research; while in the 1990s, hematology, sleep, and fractals were a focus of coherent topics. 

\section{Discussion}

In this work, we have presented a framework for generating diachronic embeddings of scholarly periodicals, enabling a systematic analysis of the evolving topics in published research over time. It allows us to identify the directions of semantic shifts by mapping periodicals onto conceptual axes and to detect emerging research topics through tracking the formation of tight clusters. Our diachronic embeddings provide new measurements that address both conceptual and computational challenges. For instance, they enable us to quantify the proximity of periodicals to various disciplines and track the changing landscape of interdisciplinary periodicals. Through comprehensive demonstrations of the utility of our diachronic embeddings, our work represents one of the first efforts to extend embedding-based approaches in the science of science in a diachronic context. 

When quantifying the extent of semantic changes, we are aware of another popular method that is based on the alignment between two vector spaces~\cite{hamilton2016diachronic}. However, we have found that semantic changes based on this approach are highly influenced by the number of periodicals in different fields: Periodicals in larger fields on average have smaller changes, which arises from the inherent setup of aligning two vector spaces (see SI). 

We acknowledge limitations in our study. First, our embeddings may be influenced by errors present in the dataset. Throughout our research, we have noticed instances of periodical entities that were incorrectly disambiguated, such as journals with the same acronym being misidentified as a single entity, or difficulties in accurately representing journals with name changes (see Table~S3). Second, we split the dataset by publication decade and trained models using citations from within the same decade, thereby excluding cross-decade citations. Alternative techniques, such as temporal network embedding, may offer a way to incorporate these citations. Third, while our diachronic embeddings assign decade-specific vectors to each periodical, a single vector may not adequately represent multiple contexts, particularly for multidisciplinary periodicals. Future research could explore the effectiveness of contextualized embeddings, such as those derived from the Transformers, in downstream tasks related to scientific evolution. However, previous research suggests that Transformers-based embedding methods do not necessarily outperform word2vec in detecting semantic changes~\cite{schlechtweg-etal-2020-semeval}. 

Despite these limitations, we demonstrate that diachronic embeddings can serve as a valuable tool for the science of science research. Future studies could explore additional dimensions, such as scientific discourse in texts, to create new embedding methods and address questions informed by the insights gained from these embeddings.

\section{Methods}

\subsection{Dataset}

We use a version of the Microsoft Academic Graph (MAG) dataset retrieved in December 2021~\cite{sinha2015overview}. MAG is a large-scale heterogeneous network that contains papers, citations, authors, journals, and more. For our analysis, we focus on journal and conference papers published from 1950 and onward, as earlier papers have scarce references. Our dataset contains $93,311,527$ papers published in $53,412$ periodicals, with a total of $554,338,274$ citations between these papers. 

Since discipline category information was not provided in MAG, we use Scopus subject area categories to label periodicals. Scopus employs the ASJC (All Science Journal Classification) scheme to categorize journals into 27 subject areas, which are further organized into four top-level subject fields: physical, life, health, and social science. For journals that belong to multiple subject areas, we select the one most commonly shared by the journal's 50 closest journals, determined by cosine similarity between their embeddings. We match $22,364$ journals between MAG and Scopus based on their names, of which $21,895$ are covered in our embeddings. 

\subsection{Model}

For each citation network between papers, we generate $N$ citation trails, $\{T_1, T_2, \cdots, T_N\}$, by performing random walks on the network. Each node serves as the starting point of a random walk for five times, to ensure that every paper is visited, and each walk randomly follows outgoing links until reaching a dead end (a paper without outgoing links). For each trail $T$, represented as a sequence of papers $( P^T_1, P^T_2, \cdots, P^T_{|T|} )$, we create a corresponding periodical trail $\gamma_T = ( V^T_1, V^T_2, \cdots, V^T_{|T|})$, where $V^T_i$ indicates the publication venue (periodical) of $P^T_i$. We then filter out periodical trails of length 1 (\emph{i.e.}, $|T| = 1$), as they do not capture citation relationships between papers, as well as periodical trails composed solely of identical periodicals. We set a minimum frequency threshold of 50 for periodicals, meaning that those with fewer than 50 occurrences are excluded from the embedding model due to data sparsity. Table~S1 provides summary statistics for each decade. 

For each decade, we use the corpus of periodical trails to train a word2vec model with the skip-gram with negative sampling (SGNS) method~\cite{mikolov2013distributed}. Based on our previous research~\cite{peng2021neural}, we set the following hyperparameters: context window size $w = 10$, embedding dimensions $D = 100$, and number of sampled negative pairs for each positive input pair $k = 5$. We also conduct validation experiments with different hyperparameter configurations (see Table~S4). After training, the ``input'' vectors from the word2vec model are used as the periodical embeddings. We utilize the Gensim package for embedding training~\cite{rehurek_lrec} and obtain embeddings for $43,476$ periodicals, after dropping $9,936$ periodicals because of data filtering.

\subsection{Data and code availability}
MAG is publicly available at \url{https://zenodo.org/records/6511057}. The code used for data analysis and generating all the results presented in this work is available at \url{https://github.com/netknowledge/diachronic-p2v} \cite{coderepo}.

\begin{acknowledgments}
This work is supported by the National Natural Science Foundation of China (72204206), City University of Hong Kong (Project No. 9610552, 7005968), and the Hong Kong Institute for Data Science.
\end{acknowledgments}

\clearpage

\setcounter{figure}{0}
\makeatletter
\renewcommand{\thefigure}{S\@arabic\c@figure}
\makeatother

\setcounter{table}{0}
\makeatletter
\renewcommand{\thetable}{S\@arabic\c@table}
\makeatother

\setcounter{section}{0}
\makeatletter
\renewcommand{\thesection}{S\@arabic\c@section}
\makeatother

\begin{center}{\Large\textbf{Supporting Information}}\end{center}

\section{Supporting Information Text}
\subsection{Quantifying semantic change of periodicals}
\subsubsection{Local neighbor perspective}

\begin{figure}[h!]
\centering
\includegraphics[trim=0 2mm 0 0, width=\linewidth]{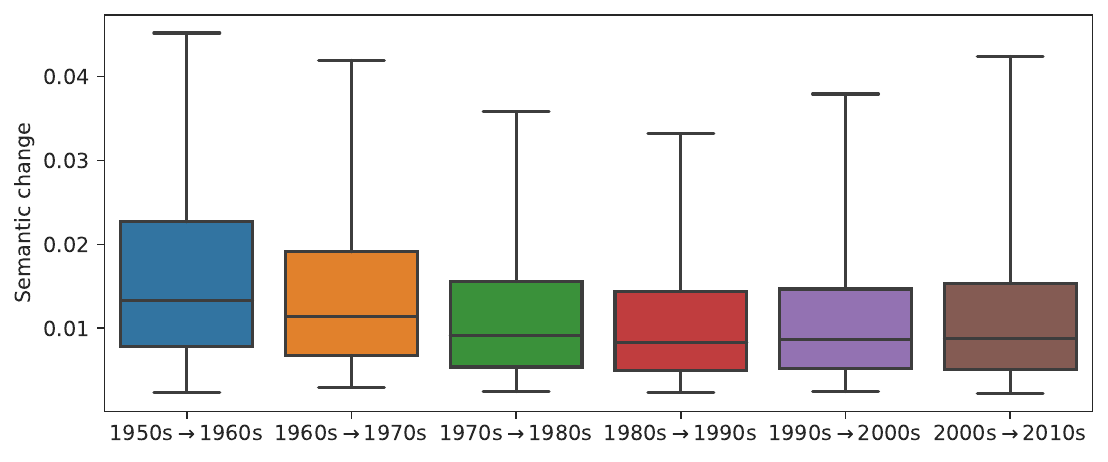}
\caption{Distributions of semantic changes based on local neighbors. Lower and upper whiskers correspond to 5th and 95th percentiles, and fliers are not shown for clarity.}
\label{fig:si:change-local-dist}
\end{figure}

\begin{figure}[h!]
\centering
\includegraphics[trim=0 2mm 0 0, width=\linewidth]{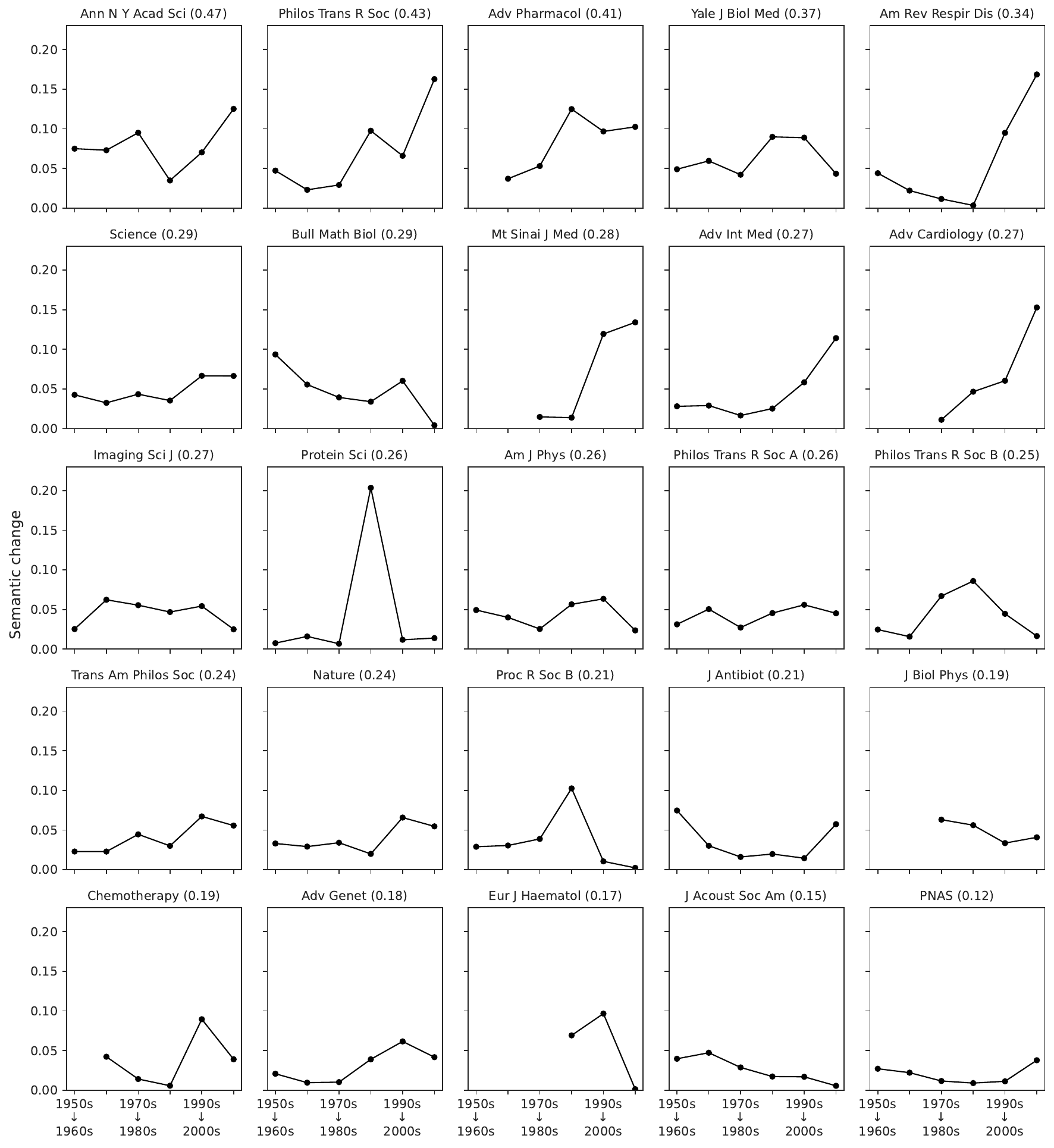}
\caption{Semantic changes of selected periodicals. Numbers in the parentheses in the titles are total changes. Semantic changes are calculated based on local neighbors.}
\label{fig:si:venue-local-dist-case-1}
\end{figure}

\begin{figure}[h!]
\centering
\includegraphics[trim=0 2mm 0 0, width=\linewidth]{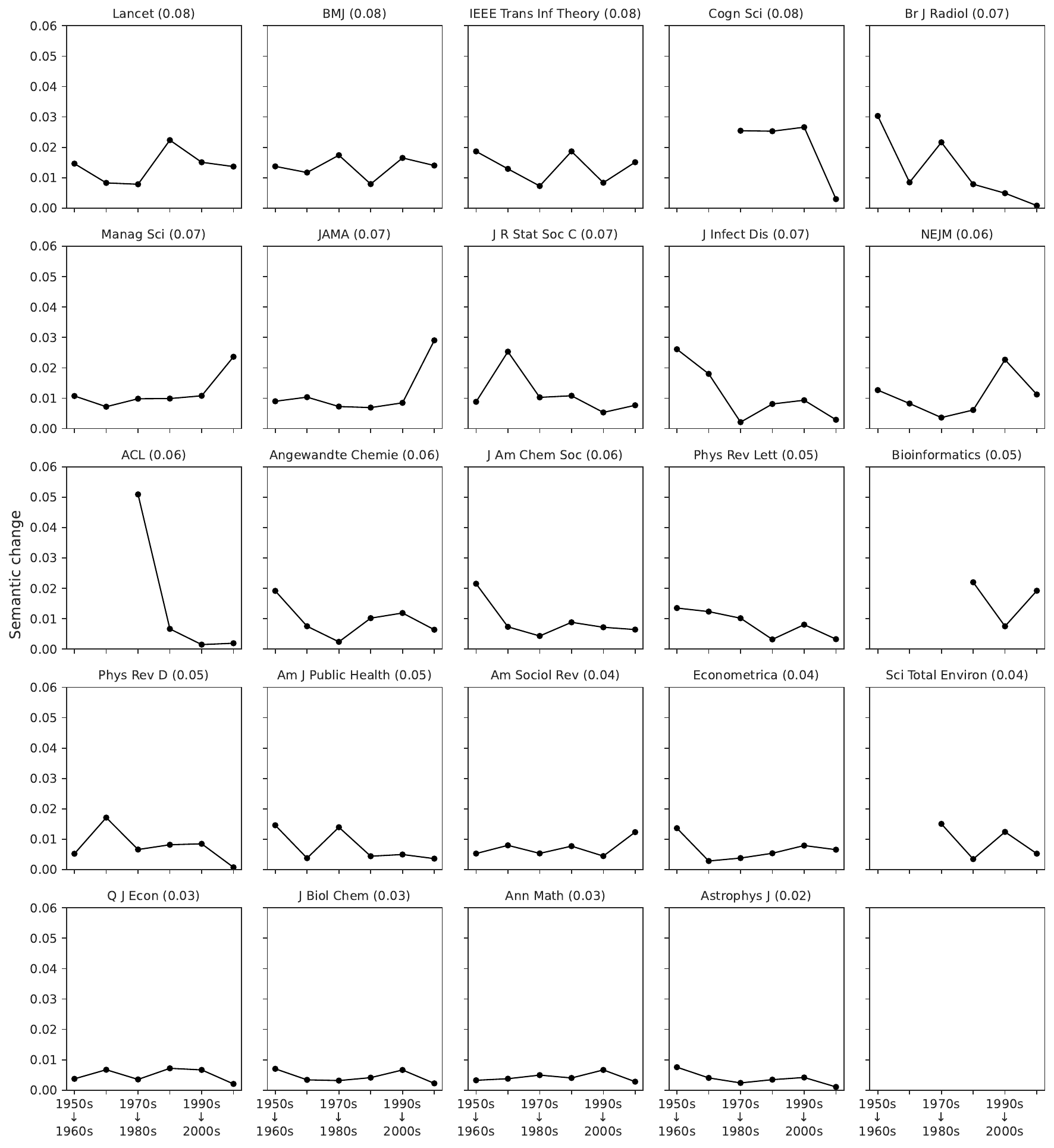}
\caption{Semantic changes of selected periodicals. Numbers in the parentheses in the titles are total changes. Semantic changes are calculated based on local neighbors.}
\label{fig:si:venue-local-dist-case-2}
\end{figure}

\begin{figure}[h!]
\centering
 \includegraphics[trim=0 3mm 0 0, width=\linewidth]{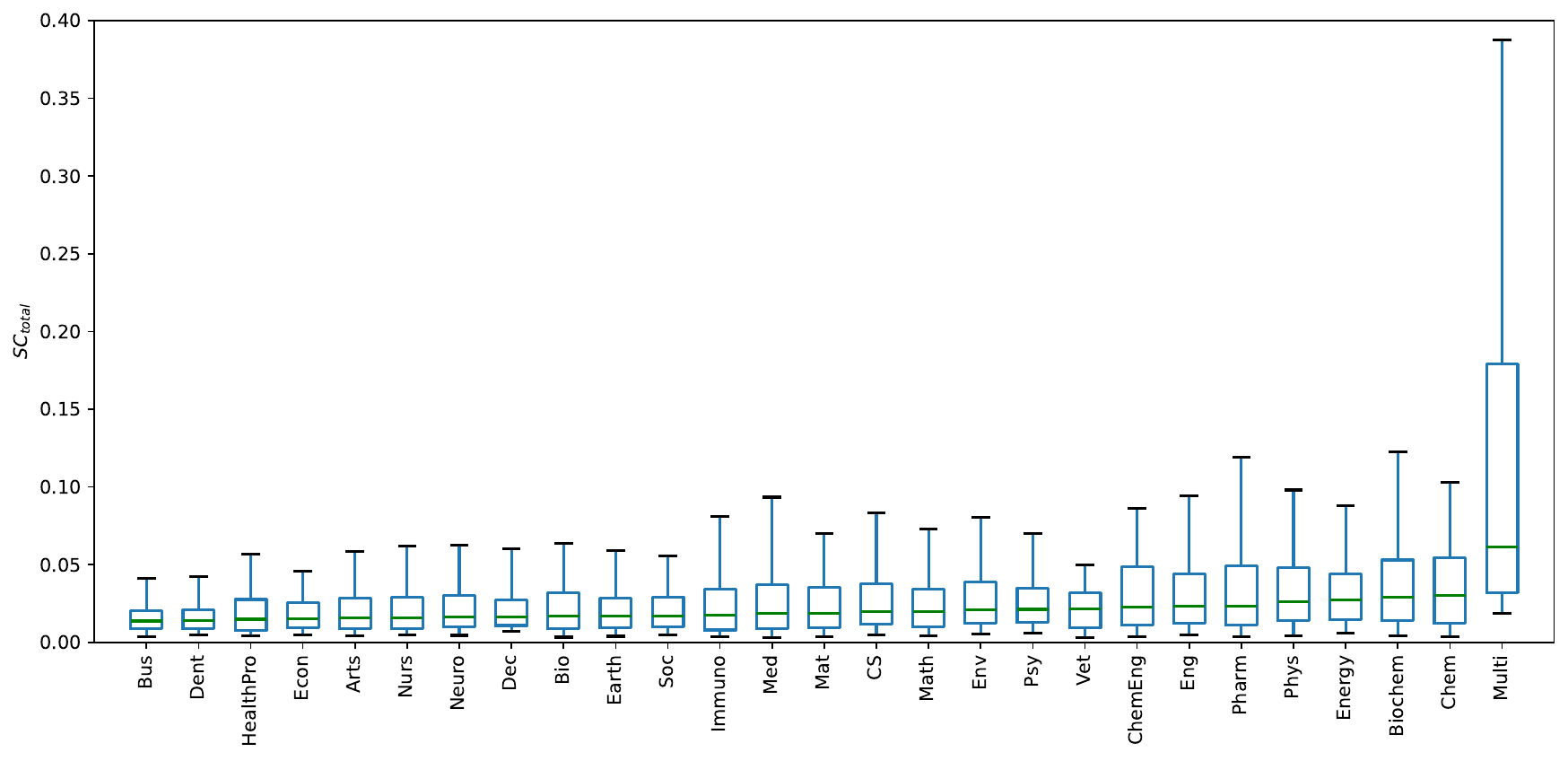}
\caption{Distributions of periodicals' total semantic changes untill the 2010s by field (using local neighbor measurement), as designated in the Scopus database. Fields are arranged from left to right based on the median.}
\label{fig:si:change-local-field}
\end{figure}

In the main text, we have quantified the semantic changes of periodical $i$ between $t_1$ and $t_2$ from its local neighbor perspective, which is the cosine distance between the two vectors representing the cosine similarities between $i$ and its nearest neighbors at $t_1$ and $t_2$. Fig.~\ref{fig:si:change-local-dist} presents the distributions of semantic changes between two consecutive decades, indicating limited changes across decades. Figs.~\ref{fig:si:venue-local-dist-case-1}--\ref{fig:si:venue-local-dist-case-2} show semantic changes of selected periodicals. Fig.~\ref{fig:si:change-local-field} shows the distributions of semantic changes by field, suggesting that periodicals belonging to natural science disciplines, including Chemistry, Biochemistry, and Energy, as well as Multidisciplinary periodicals, tend to have greater semantic changes, compared to those peers that belong to Humanities, Social Sciences, and Business. 

\subsubsection{Global alignment perspective}

We also explore another quantification of semantic change, which is based on global alignment between two vector spaces at $t_1$ and $t_2$, given that they may correspond to different coordinate systems. Specifically, let $\mathbf{V}^{t} \in \mathbb{R}^{d \times |v|}$ denote the matrix of embedding vectors learned at time $t$ and $d$ is the embedding dimension. The alignment is to find the best rotational operation that most closely maps $\mathbf{V}^{t_1}$ to $\mathbf{V}^{t_2}$ for the shared set of periodicals at $t_1$ and $t_2$. Formally, the alignment is solved through orthogonal Procrustes analysis: 
\begin{equation} \label{eq:si:align}
\mathbf{R}^{(t_1 \rightarrow t_2)} = \mathop{\arg\min}\limits_{\mathbf{Q}^\top\mathbf{Q}=\mathbf{I}} \left\Vert \mathbf{QV}^{t_1} - \mathbf{V}^{t_2}\right\Vert_F \, ,
\end{equation}
where $\left\Vert \cdot \right\Vert_F$ is the Frobenius norm of a matrix and $\mathbf{R}^{(t_1 \rightarrow t_2)} \in \mathbb{R}^{d \times d}$ is the identified rotational operation. Eq.~\ref{eq:si:align} can be solved using the application of SVD~\cite{schonemann1966procrustes}. Then, the semantic change of periodical $i$ is the cosine distance between $v_i^{t_2}$ and $\mathbf{R}^{(t_1 \rightarrow t_2)} v_i^{t_1}$, the aligned vector of $v_i^{t_1}$ in the vector space at $t_2$. Here we set $t_1$ and $t_2$ to be two consecutive decades. 

\begin{figure}[h!]
\centering
\includegraphics[trim=0 2mm 0 0, width=\linewidth]{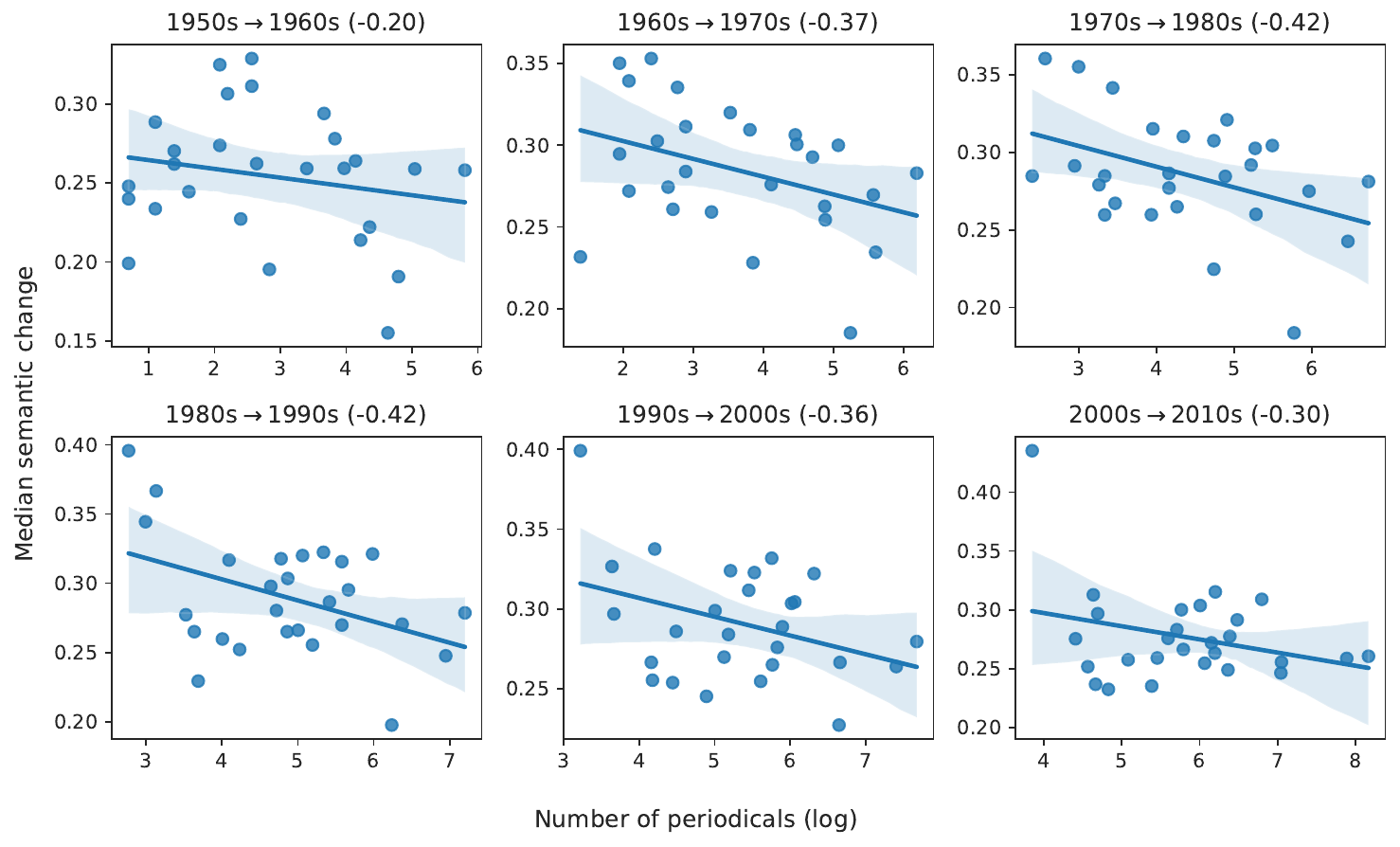}
\caption{Disciplines with more periodicals tend to experience less semantic changes, which are calculated based on global alignment. Numbers in the parentheses in the titles are correlation coefficients.}
\label{fig:si:field-size-change}
\end{figure}

However, we stress that although this method has been used in previous studies to detect semantic changes of words~\cite{hamilton2016diachronic}, it has unequal effects for periodicals from different disciplines. Specifically, the goal of Eq.~\ref{eq:si:align} is to find the best rotation such that the \emph{sum} of the vector differences across periodicals achieves minimum. Therefore, disciplines with more periodicals, such as Medicine, may play a larger role in determining the alignment matrix, and consequently those periodicals may have smaller semantic changes. Fig.~\ref{fig:si:field-size-change} empirically demonstrates this discipline size effect, showing that disciplines with more periodicals tend to experience less semantic changes. 

\begin{figure}[h!]
\centering
\includegraphics[trim=0 2mm 0 0, width=\linewidth]{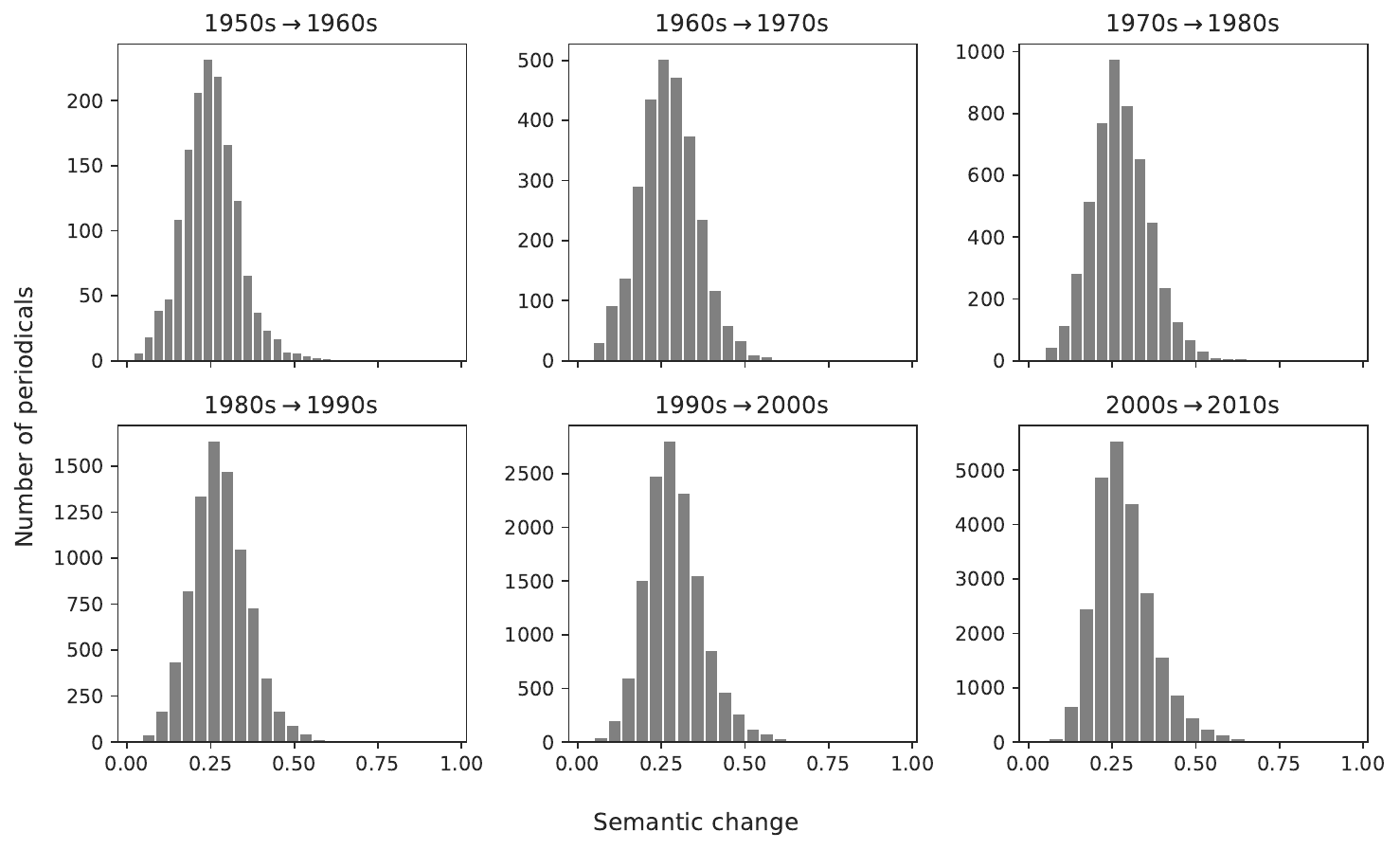}
\caption{Histograms of global alignment based semantic changes.}
\label{fig:si:jnl-aligned-dist-hist}
\end{figure}

\begin{figure}[h!]
\centering
\includegraphics[trim=0 4mm 0 0, width=\linewidth]{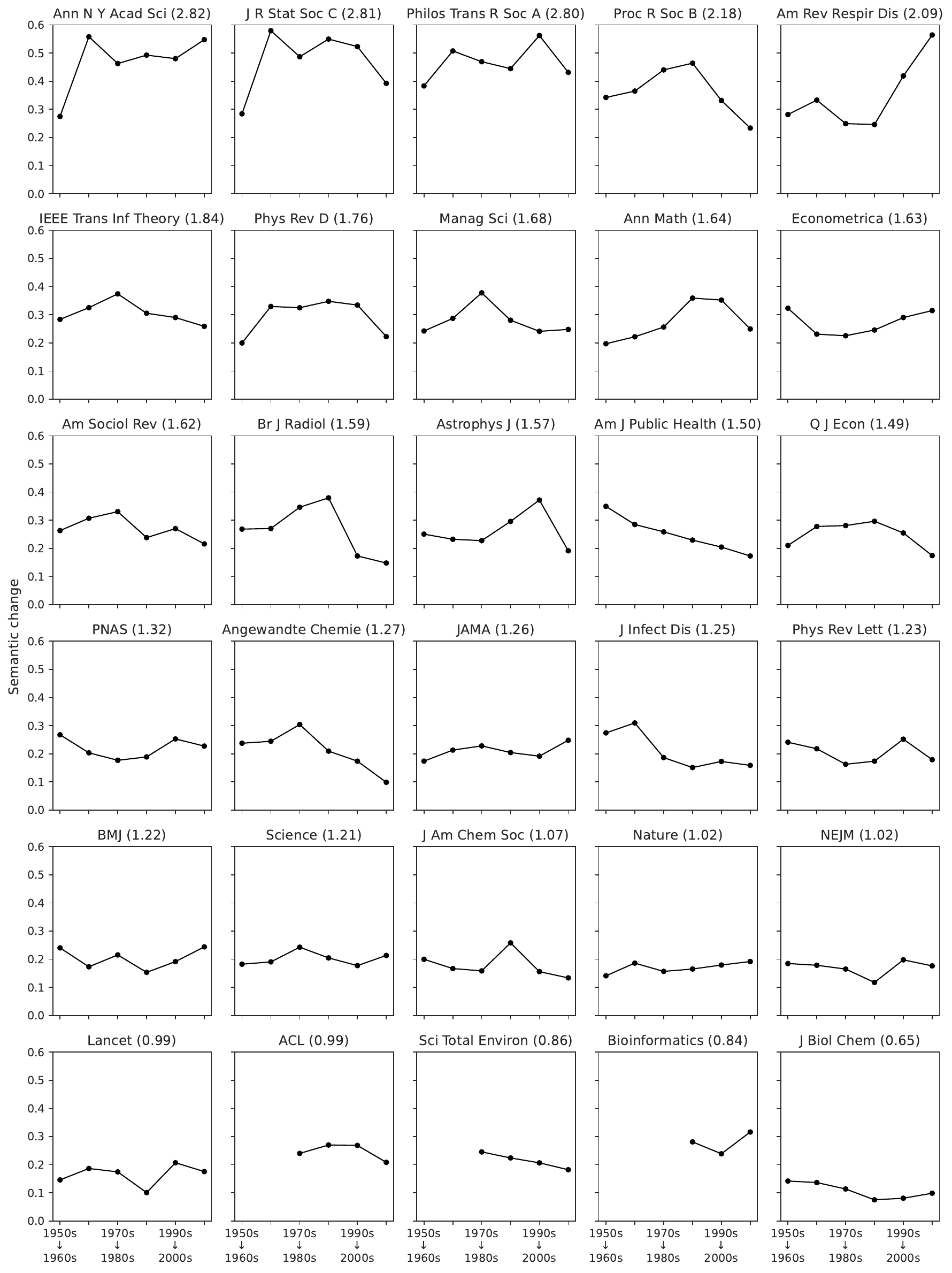}
\caption{Semantic changes of selected periodicals. Numbers in the parentheses in the titles are total changes. Semantic changes are calculated based on global alignment.}
\label{fig:si:venue-aligned-dist-cases}
\end{figure}

\begin{figure}[h!]
\centering
\includegraphics[trim=0 4mm 0 0, width=1\linewidth]{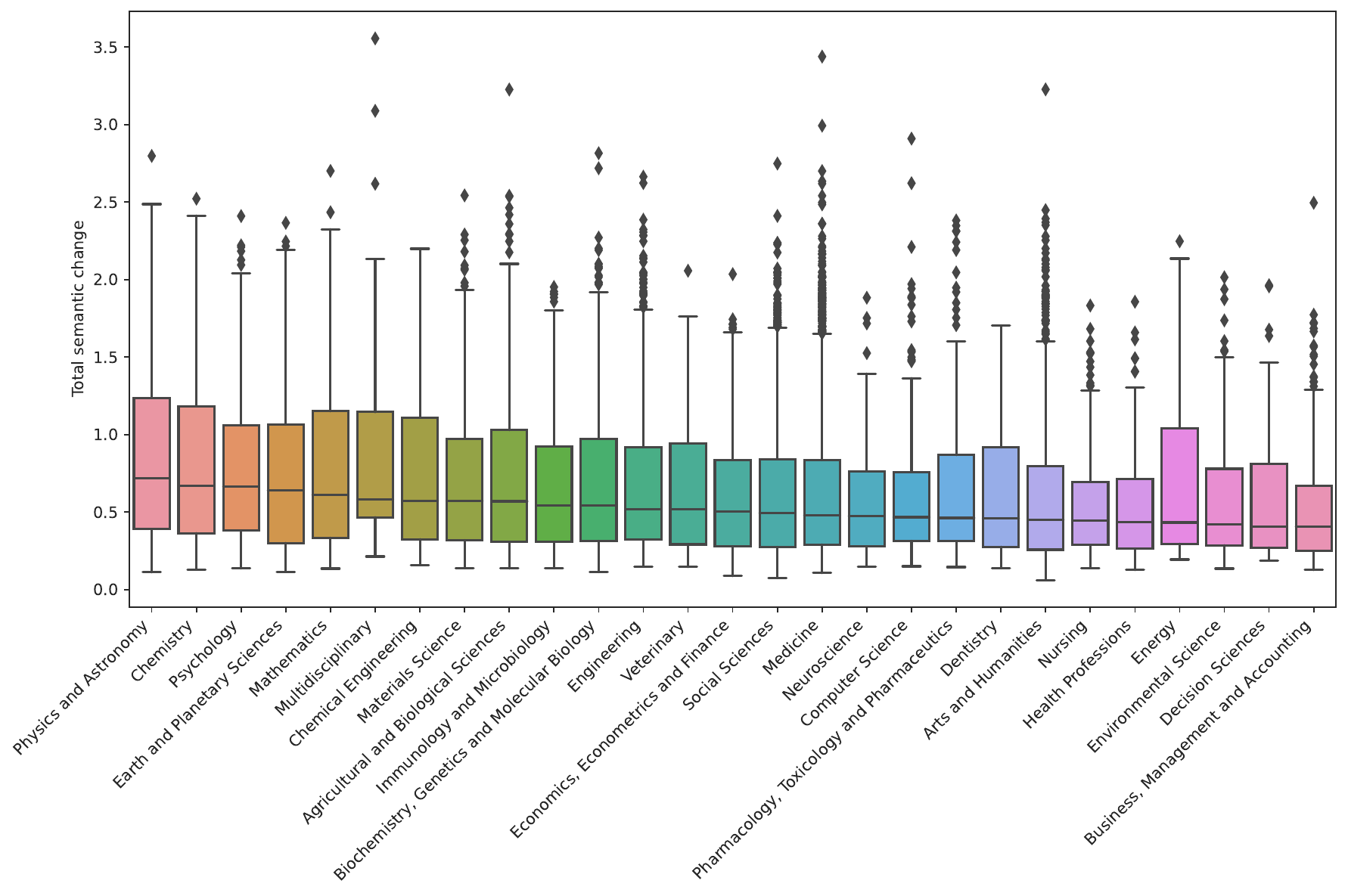}
\caption{Distributions of total semantic changes of periodicals by field. Fields are arranged from left to right based on the decreasing order of the median.}
\label{fig:si:venue-aligned-dist-field}
\end{figure}

Bearing this caveat in mind, we nevertheless proceed to present the results about semantic changes of periodicals from the global alignment perspective. Fig.~\ref{fig:si:jnl-aligned-dist-hist} indicates that across time, periodicals have limited semantic changes. Fig.~\ref{fig:si:venue-aligned-dist-cases} plots semantic changes of a set of selected periodicals, suggesting that \emph{Annals of Mathematics} and \emph{American Sociological Review} have larger changes than \emph{Nature}, whereas the opposite is observed when semantic changes are measured using local neighbors. Finally, Fig.~\ref{fig:si:venue-aligned-dist-field}, which shows the distributions of total semantic changes by field, indicates that Physics and Astronomy, Chemistry, and Psychology periodicals are more vibrant than those from biomedical and health fields, reinforcing partially that semantic changes are dependent on field size. 

\clearpage

\section{Supporting Information Figures}

\begin{figure}[h!]
\centering
\includegraphics[trim=0 4mm 0 0, width=\linewidth]{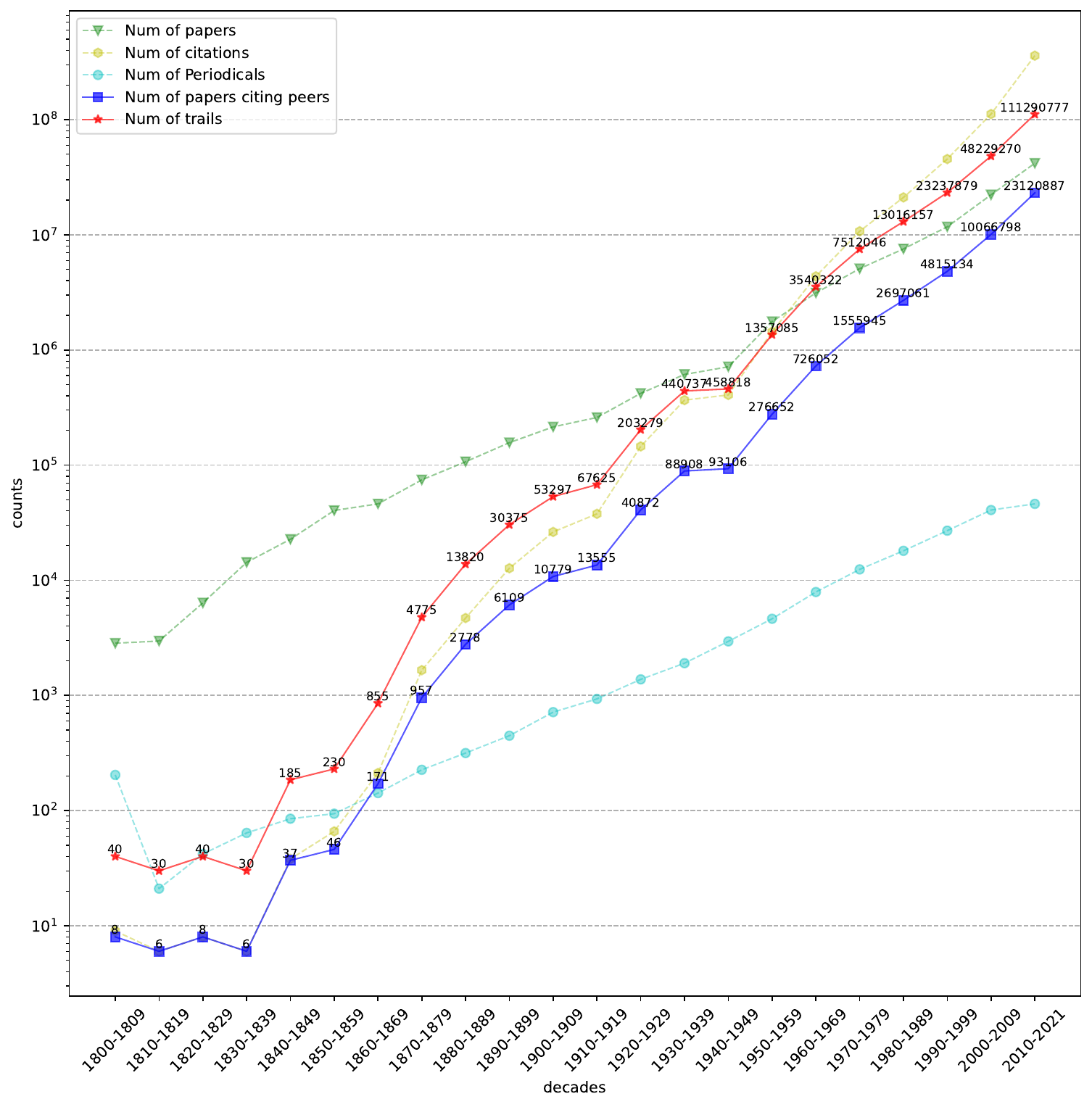}
\caption{Summary statistics by decade. Most of these statistics grow exponentially, and the number of citations among papers in the same decade shows the most significant increment rate after the 1950s. The number of generated trails has been maintained at about five times the number of papers citing their peers.}
\label{fig:si:summary}
\end{figure}

\begin{figure}[h!]
\centering
\includegraphics[trim=0 0mm 0 0, width=1\linewidth]{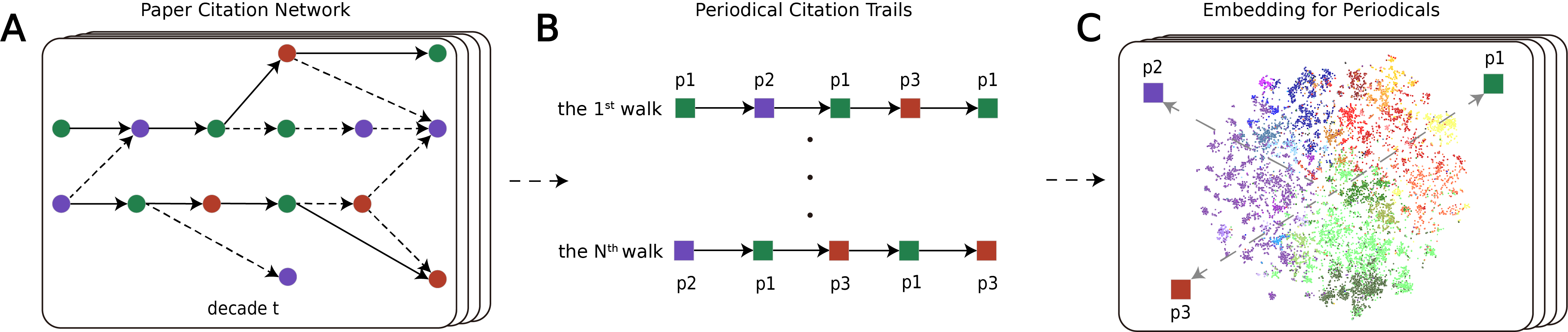}
\caption{A schematic illustration of obtaining diachronic periodical embeddings. 
(A) For each decade, we build a paper citation network from the MAG dataset representing citation relationships between papers published in the decade. 
(B) For each citation network, we perform random walks, by recursively setting every paper as the starting point of the random walk, and randomly choose the next point following the citation flow until we reach a dead end. We then map the sequences of visited papers to the sequences of periodicals, which are our corpora of ``sentences''. 
(C) For each corpus, we use word2vec to generate embedding for periodicals that occurred in trails using the skip-gram with negative sampling (SGNS) method, with $D=100, W=10$. A 2-D projection (obtained by applying t-SNE\cite{maaten2008visualizing}) of overall journal vectors is presented, where each dot represents a journal, and its color denotes its discipline designated in the Scopus ASJC (All Science Journal Classification) scheme (multidisciplinary journals are colored in black). This example is generated using data from the 2010s. By repeating A-C for corpora obtained over 7 decades, from the 1950s to the 2010s, diachronic periodical embeddings could be generated. It can be observed that the embedding space is being overpopulated (Figs.~\ref{fig:si:map_of_sci_1950s}--\ref{fig:si:map_of_sci_2010s} show the 2-d projections of periodical embeddings in the other decades), as a result of increasing number of periodicals (see Table~\ref{fig:si:summary}).}
\label{fig:si:schema}
\end{figure}

\begin{figure}[h!]
\centering
\includegraphics[width=1\linewidth]{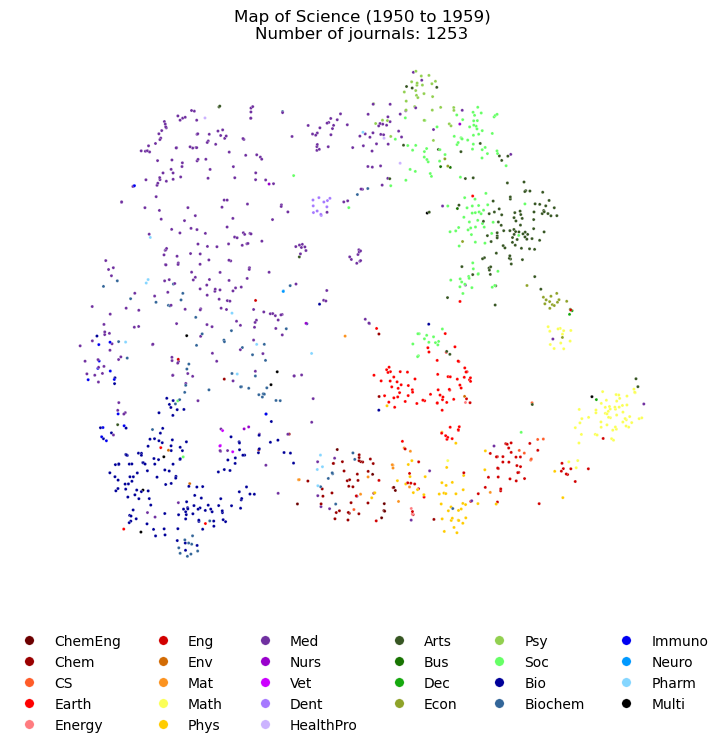}
\caption{2-D projection of journal embeddings using data from the 1950s.}
\label{fig:si:map_of_sci_1950s}
\end{figure}

\begin{figure}[h!]
\centering
\includegraphics[width=1\linewidth]{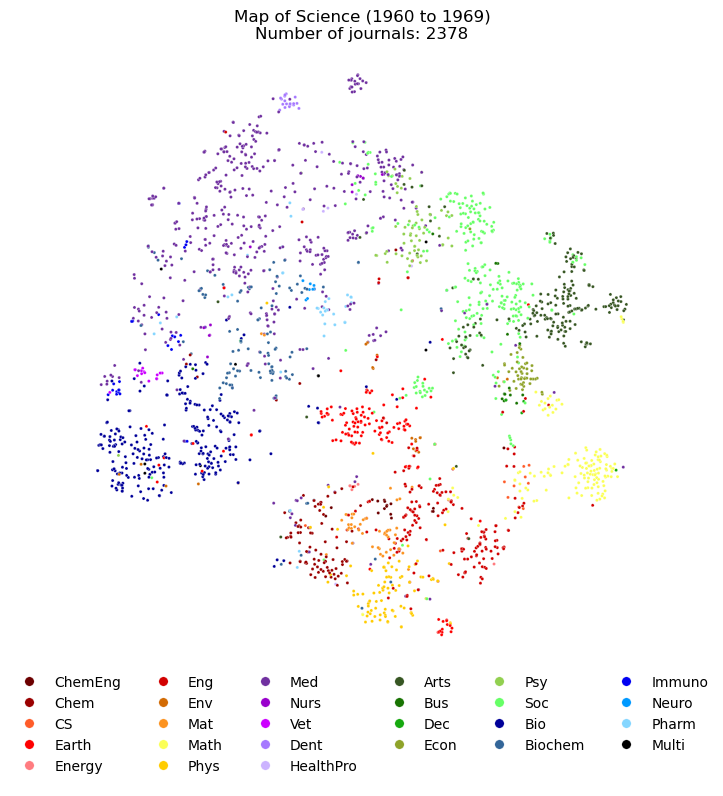}
\caption{2-D projection of journal embeddings using data from the 1960s.}
\label{fig:si:map_of_sci_1960s}
\end{figure}

\begin{figure}[h!]
\centering
\includegraphics[width=1\linewidth]{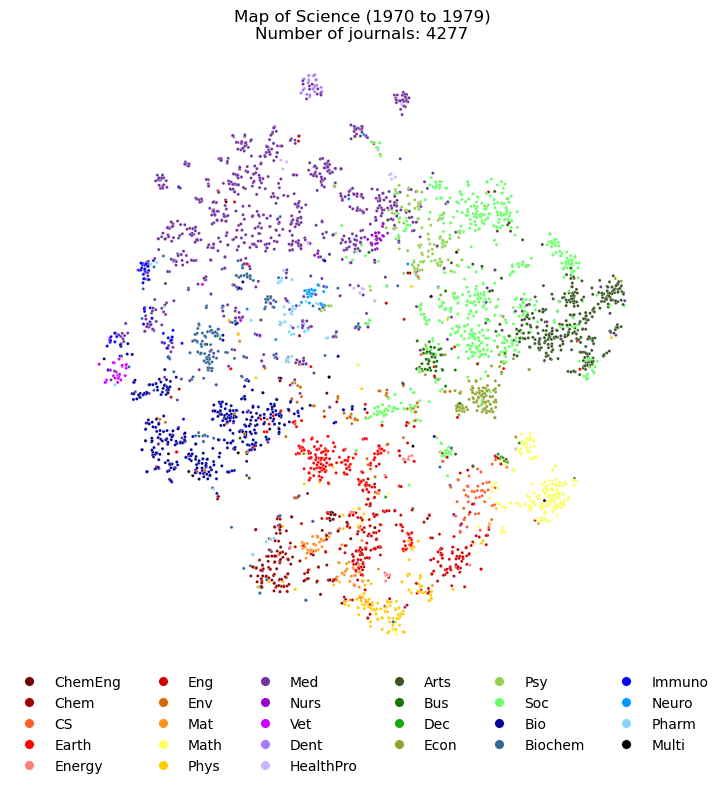}
\caption{2-D projection of journal embeddings using data from the 1970s.}
\label{fig:si:map_of_sci_1970s}
\end{figure}

\begin{figure}[h!]
\centering
\includegraphics[width=1\linewidth]{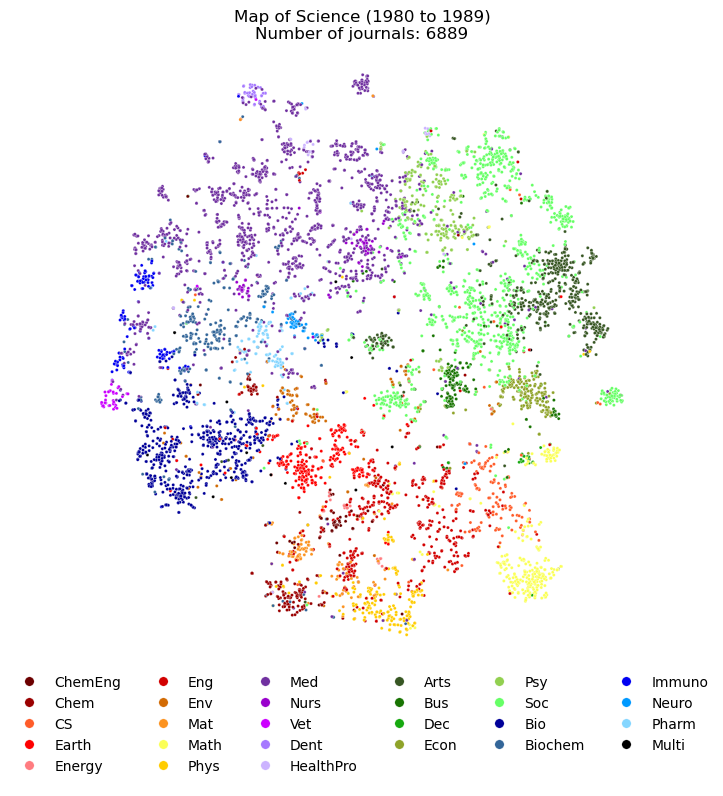}
\caption{2-D projection of journal embeddings using date from the 1980s.}
\label{fig:si:map_of_sci_1980s}
\end{figure}

\begin{figure}[h!]
\centering
\includegraphics[width=1\linewidth]{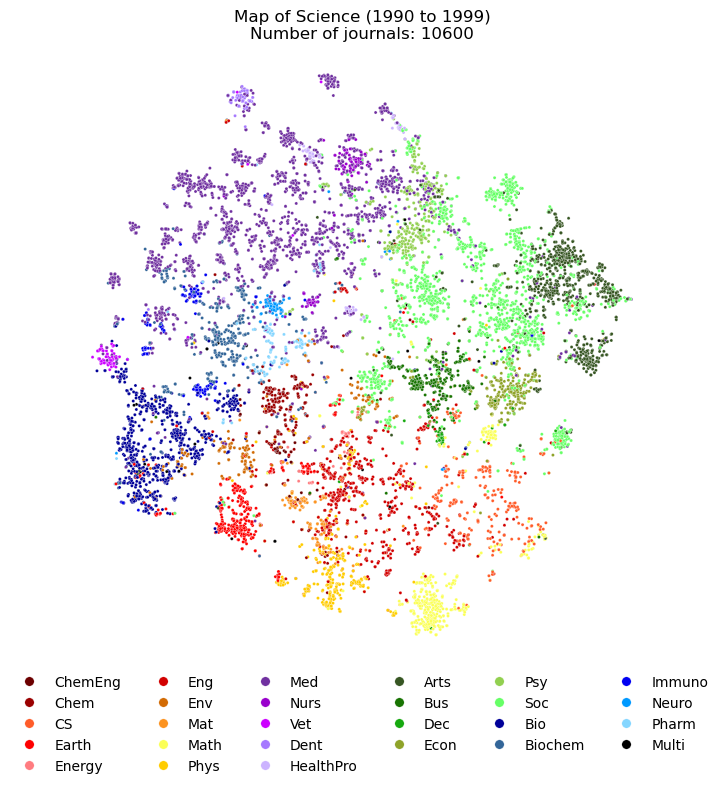}
\caption{2-D projection of journal embeddings using date from the 1990s.}
\label{fig:si:map_of_sci_1990s}
\end{figure}

\begin{figure}[h!]
\centering
\includegraphics[width=1\linewidth]{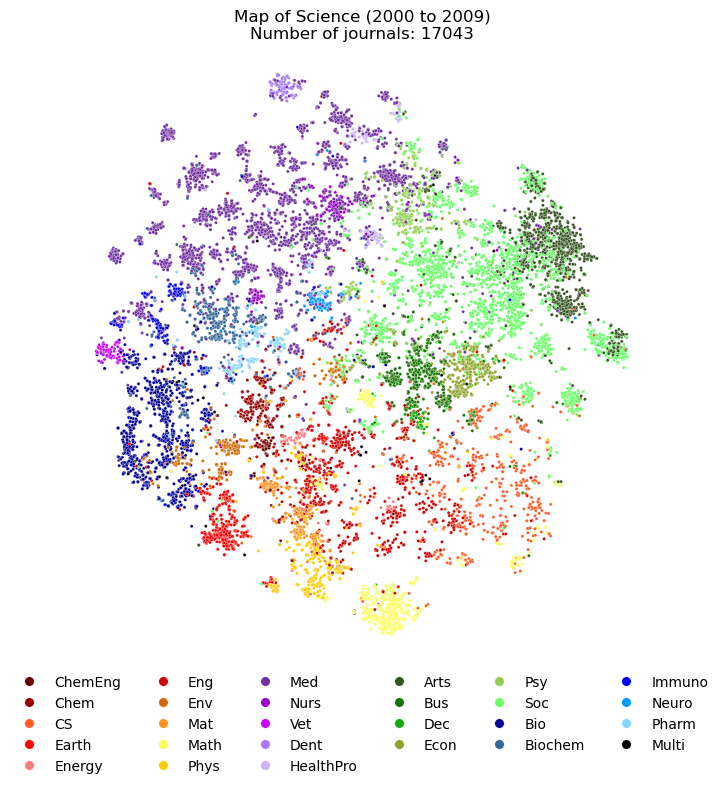}
\caption{2-D projection of journal embeddings using data from the 2000s.}
\label{fig:si:map_of_sci_2000s}
\end{figure}

\begin{figure}[h!]
\centering
\includegraphics[width=1\linewidth]{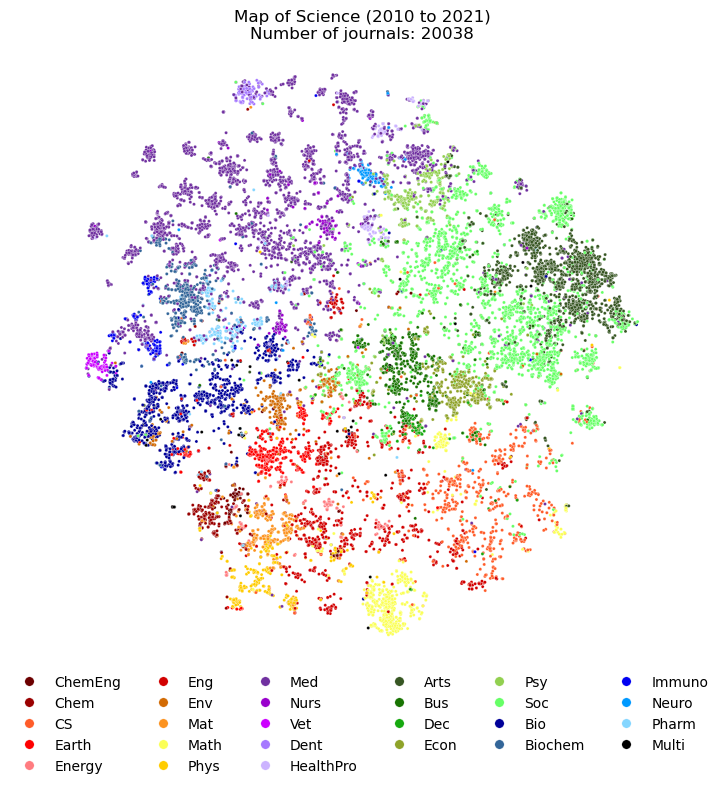}
\caption{2-D projection of journal embeddings using data from the 2010s.}
\label{fig:si:map_of_sci_2010s}
\end{figure}

\begin{figure}[h!]
\centering
\includegraphics[trim=0 2mm 0 0, width=.7\linewidth]{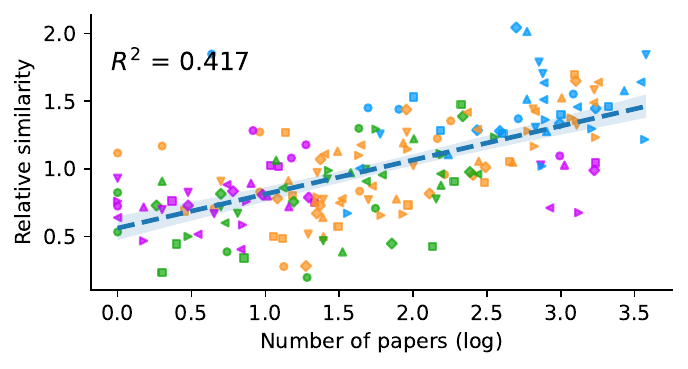}
\caption{Validating diachronic periodical embeddings using \emph{Science}.}
\label{fig:si:validation_science}
\end{figure}

\clearpage

\begin{figure}[h!]
\centering
\includegraphics[trim=0 2mm 0 0, width=.7\linewidth]{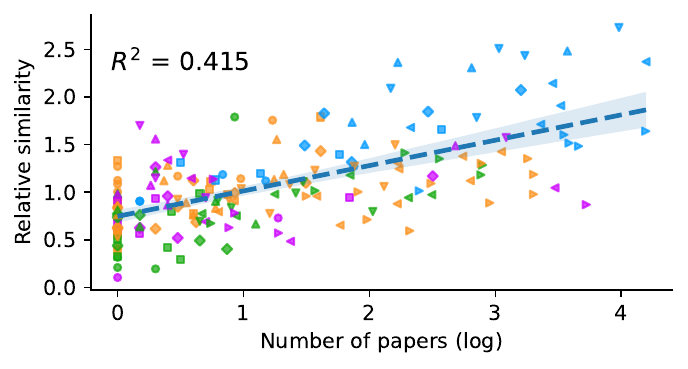}
\caption{Validating diachronic periodical embeddings using \emph{PNAS}.}
\label{fig:si:validation_pnas}
\end{figure}

\begin{figure}[h!]
\centering
\includegraphics[trim=0 5mm 0 0, width=\linewidth]{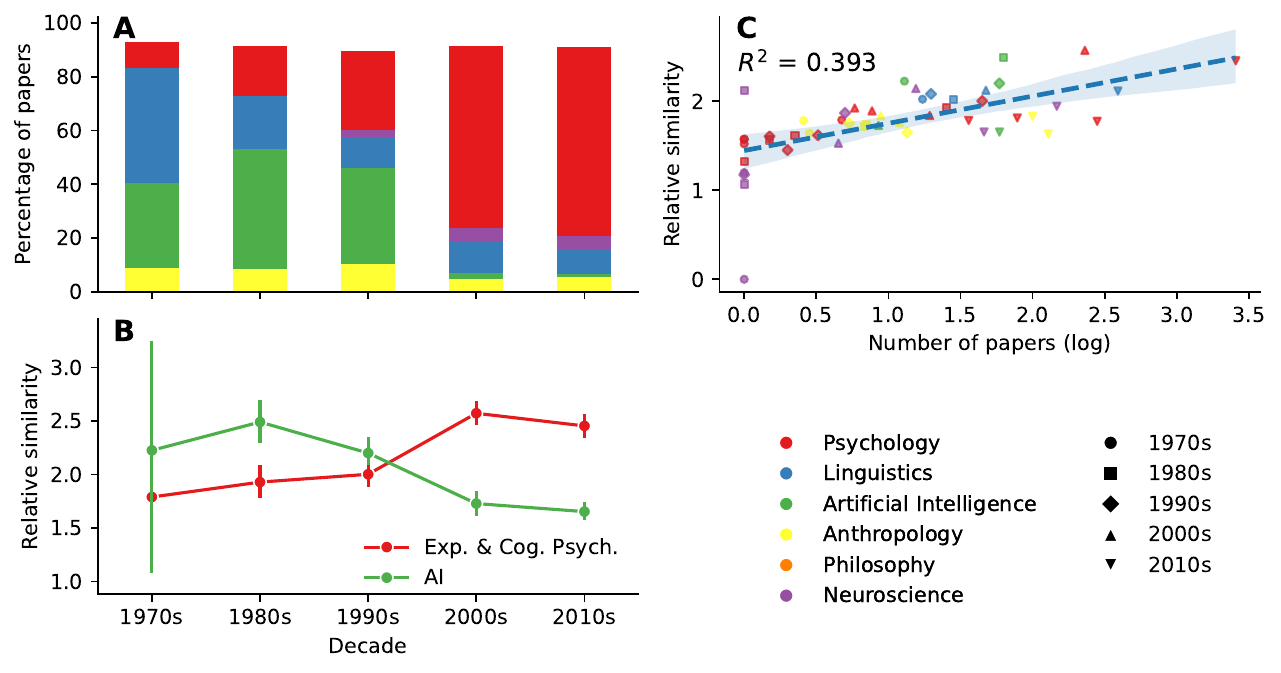}
\caption{Validating our diachronic periodical embeddings using \emph{Cognitive Science}. 
(A) Percentage of papers in 6 founding disciplines (defined in \cite{nunez2019happened}) by decade. Papers in the 2010s refer to those published in 2010--2021 for simplicity. 
(B) Relative similarity between \emph{Cognitive Science} and periodicals from the 2 focused ASJC categories. Relative similarity is defined as the average cosine similarity between \emph{Cognitive Science} and all periodicals belonging to that ASJC category, divided by the average cosine similarity between \emph{Cognitive Science} and all periodicals. 
(C) The correspondence between publication volume and similarity. Color represents founding disciplines and the shape of point marks decade.}
\label{fig:si:validation_cognitive_science}
\end{figure}

\begin{figure}[h!]
\centering
\includegraphics[trim=0 5mm 0 0, width=\linewidth]{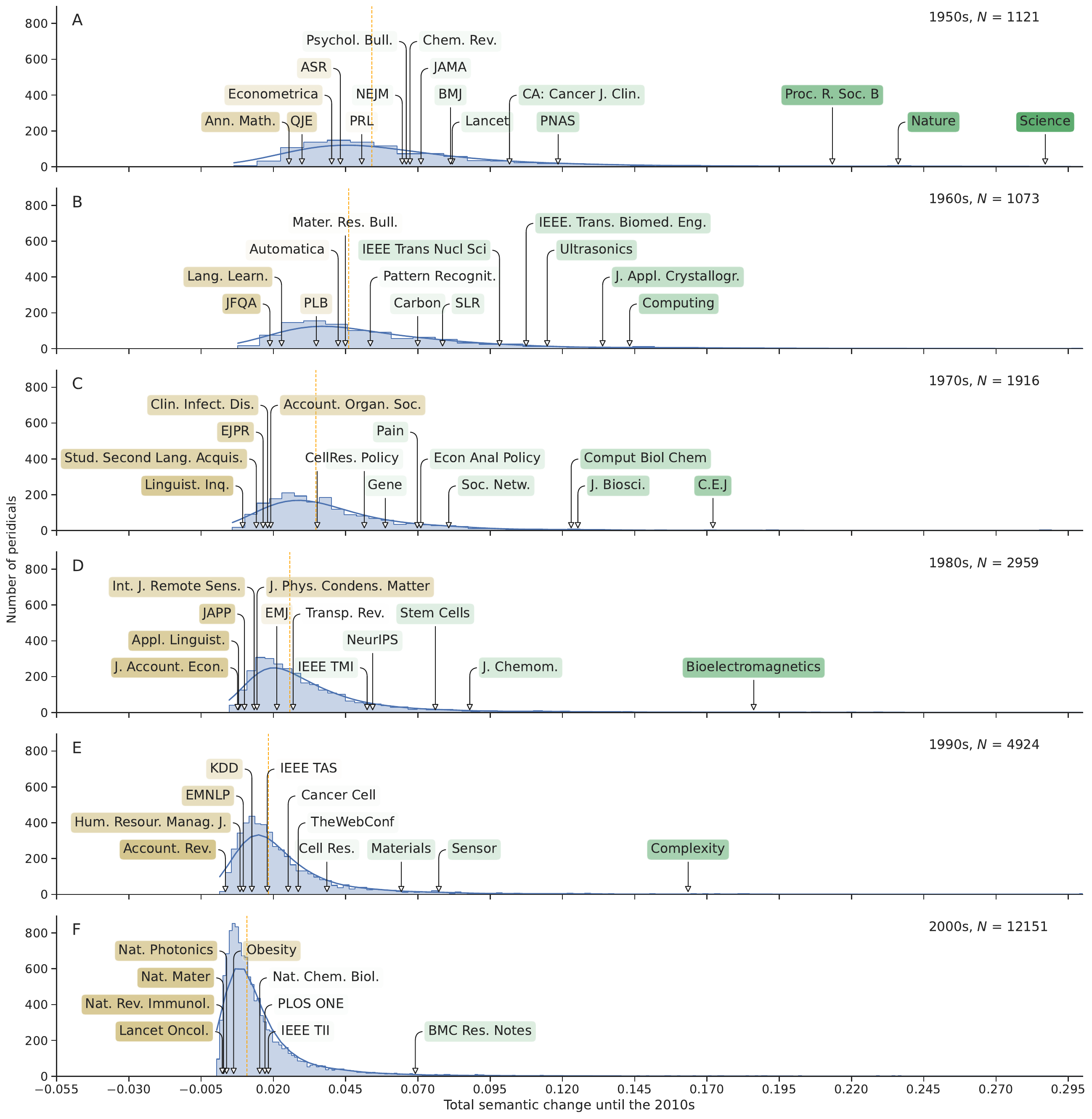}
\caption{Distributions of total semantic changes of periodicals. We group periodicals based on the decades when they were established and show the distributions for each group. Dashed vertical lines mark the medians. The number of periodicals $N$ are marked on the top right on each panel. Table~\ref{tab:si:abbr} lists periodical name abbreviations.}
\label{fig:si:total_local_semantic_change}
\end{figure}

\begin{sidewaysfigure}[p]
\centering
\includegraphics[trim=15mm 0 15mm 0, width=\linewidth]{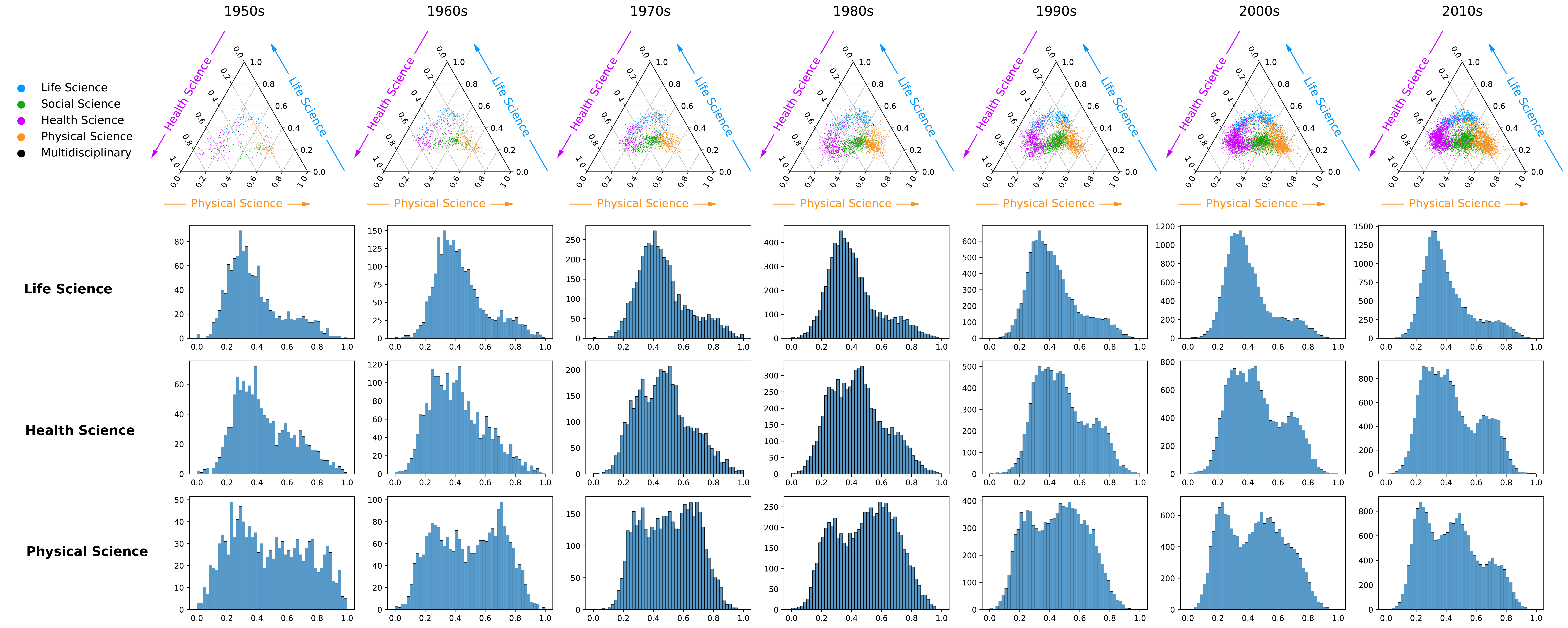}
\caption{Overall distributions of periodicals' embedding exhibited in ternary plots over 7 decades. Each dot represents a journal and is colored by the research area it belongs to.}
\label{fig:si:ternary-plot-decades}
\end{sidewaysfigure}

\begin{figure}[h!]
\centering
\includegraphics[trim=0 4mm 0 0, width=\linewidth]{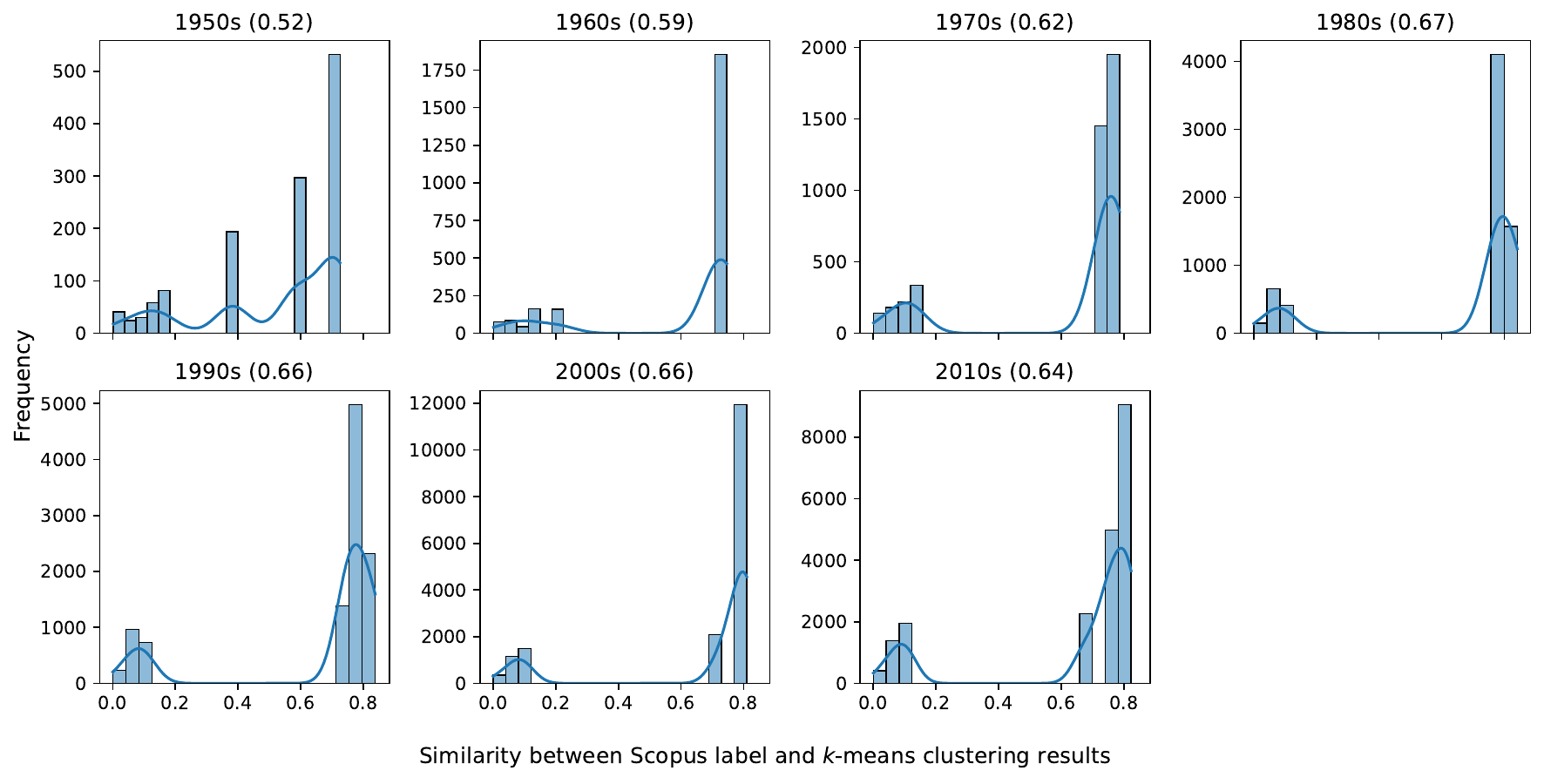}
\caption{Distributions of periodicals' similarity between their Scopus label and $k$-means results.}
\label{fig:si:sim_score_distribution}
\end{figure}

\begin{figure}[h!]
\centering
\includegraphics[trim=0 4mm 0 0, width=0.8\linewidth]{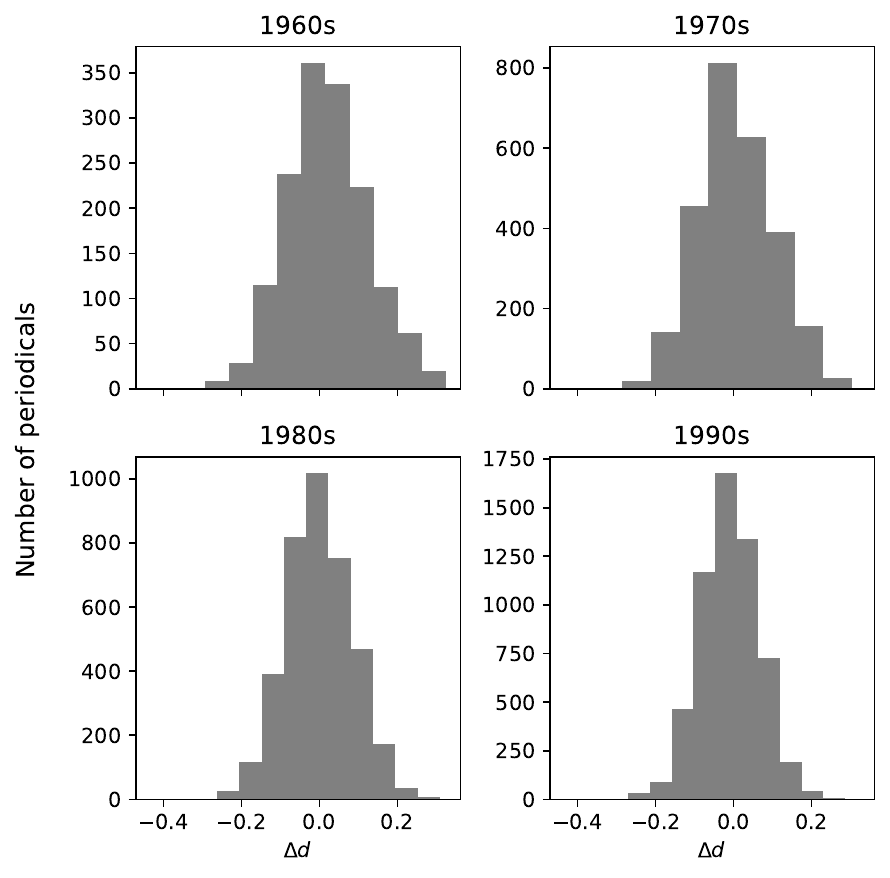}
\caption{Distribution of $\Delta d$ for periodicals established in each decade. Only a limited number of periodicals exhibit noticeable changes in distance to their $10^{th}$ nearest neighbor, compared to the 2010s.}
\label{fig:si:10-nn-dist-delta}
\end{figure}

\clearpage
\section{Supporting Information Tables}

\begin{table}[h!]
\centering
\caption{Summary statistics by decade.}
\label{tab:si:dataset-decade}
\begin{tabular}{c | r | r | r | r | r}
\toprule
Period & Papers & Citations & Papers citing peers & Walks & Periodicals  \\ 
\midrule
1950--1959 & 1764551  & 1438364      & 276652  & 1357085   & 4632            \\
1960--1969 & 3127165  & 4356908      & 726052  & 3540322   & 7907            \\
1970--1979 & 5081708  & 10729574     & 1555945 & 7512046   & 12430           \\
1980--1989 & 7541503  & 21121496     & 2697061 & 13016157  & 18023           \\
1990--1999 & 11724313 & 45487363     & 4815134 & 23237879  & 26957           \\
2000--2009 & 22251359 & 111762233    & 10066798& 48229270  & 40738           \\
2010--2021 & 41820928 & 359442336    & 23120887& 111290777 & 46074           \\ 
\bottomrule
\end{tabular}
\end{table}

\begin{table}[h!]
\centering
\caption{Number of journals in the 27 categories defined in Scopus.}
\label{tab:si:discipline distribution}
\resizebox{\linewidth}{!}{%
\begin{tabular}{l | c c | c c | c c | c c | c c | c c | c c} 
\toprule
& \multicolumn{2}{c|}{1950-1959} & \multicolumn{2}{c|}{1960-1969} & \multicolumn{2}{c|}{1970-1979} & \multicolumn{2}{c|}{1980-1989} & \multicolumn{2}{c|}{1990-1999} & \multicolumn{2}{c|}{2000-2009} & \multicolumn{2}{c}{2010-2021} \\ 
\cline{2-15}
& \multicolumn{1}{c}{count} & \multicolumn{1}{c|}{\%} & \multicolumn{1}{c}{count} & \multicolumn{1}{c|}{\%} & \multicolumn{1}{c}{count} & \multicolumn{1}{c|}{\%} & \multicolumn{1}{c}{count} & \multicolumn{1}{c|}{\%} & \multicolumn{1}{c}{count} & \multicolumn{1}{c|}{\%} & \multicolumn{1}{c}{count} & \multicolumn{1}{c|}{\%} & \multicolumn{1}{c}{count} & \multicolumn{1}{c}{\%} \\ 
\midrule
Multidisplinary & 8 & 0.63 & 9 & 0.37 & 14 & 0.32 & 16 & 0.23 & 26 & 0.24 & 46 & 0.27 & 58 & 0.29 \\
Biochemistry, Genetics and Molecular Biology & 67 & 5.23 & 112 & 4.61 & 184 & 4.25 & 306 & 4.40 & 457 & 4.28 & 662 & 3.87 & 724 & 3.61 \\
Physics and Astronomy & 48 & 3.75 & 94 & 3.87 & 151 & 3.49 & 224 & 3.22 & 330 & 3.09 & 442 & 2.59 & 459 & 2.29 \\
Chemistry & 38 & 2.97 & 91 & 3.74 & 126 & 2.91 & 167 & 2.40 & 253 & 2.37 & 329 & 1.93 & 366 & 1.83 \\
Medicine & 351 & 27.42 & 526 & 21.65 & 901 & 20.82 & 1445 & 20.79 & 2311 & 21.66 & 3921 & 22.95 & 4635 & 23.13 \\
Immunology and Microbiology & 18 & 1.41 & 26 & 1.07 & 70 & 1.62 & 135 & 1.94 & 190 & 1.78 & 261 & 1.53 & 279 & 1.39 \\
Engineering & 62 & 4.84 & 187 & 7.70 & 287 & 6.63 & 462 & 6.65 & 644 & 6.04 & 945 & 5.53 & 1146 & 5.72 \\
Arts and Humanities & 111 & 8.67 & 196 & 8.07 & 359 & 8.29 & 547 & 7.87 & 797 & 7.47 & 1203 & 7.04 & 1571 & 7.84 \\
Agricultural and Biological Sciences & 161 & 12.58 & 274 & 11.28 & 402 & 9.29 & 608 & 8.75 & 804 & 7.54 & 1215 & 7.11 & 1373 & 6.85 \\
Dentistry & 10 & 0.78 & 18 & 0.74 & 26 & 0.60 & 43 & 0.62 & 69 & 0.65 & 111 & 0.65 & 139 & 0.69 \\
Materials Science & 11 & 0.86 & 50 & 2.06 & 77 & 1.78 & 120 & 1.73 & 181 & 1.70 & 353 & 2.07 & 443 & 2.21 \\
Nursing & 5 & 0.39 & 13 & 0.53 & 30 & 0.69 & 74 & 1.06 & 142 & 1.33 & 234 & 1.37 & 263 & 1.31 \\
Psychology & 33 & 2.58 & 62 & 2.55 & 137 & 3.17 & 227 & 3.27 & 360 & 3.37 & 485 & 2.84 & 520 & 2.59 \\
Earth and Planetary Sciences & 80 & 6.25 & 139 & 5.72 & 211 & 4.88 & 279 & 4.01 & 366 & 3.43 & 514 & 3.01 & 570 & 2.84 \\
Decision Sciences & 4 & 0.31 & 2 & 0.08 & 11 & 0.25 & 19 & 0.27 & 37 & 0.35 & 78 & 0.46 & 86 & 0.43 \\
Mathematics & 73 & 5.70 & 137 & 5.64 & 200 & 4.62 & 280 & 4.03 & 437 & 4.10 & 679 & 3.97 & 744 & 3.71 \\
Pharmacology, Toxicology and Pharmaceutics & 13 & 1.02 & 35 & 1.44 & 67 & 1.55 & 118 & 1.70 & 155 & 1.45 & 308 & 1.80 & 327 & 1.63 \\
Environmental Science & 4 & 0.31 & 21 & 0.86 & 68 & 1.57 & 122 & 1.76 & 195 & 1.83 & 347 & 2.03 & 424 & 2.12 \\
Social Sciences & 132 & 10.31 & 303 & 12.47 & 651 & 15.04 & 1071 & 15.41 & 1681 & 15.76 & 2747 & 16.08 & 3250 & 16.22 \\
Neuroscience & 2 & 0.16 & 8 & 0.33 & 35 & 0.81 & 55 & 0.79 & 91 & 0.85 & 139 & 0.81 & 158 & 0.79 \\
Chemical Engineering & 8 & 0.63 & 16 & 0.66 & 28 & 0.65 & 60 & 0.86 & 72 & 0.67 & 124 & 0.73 & 128 & 0.64 \\
Veterinary & 4 & 0.31 & 15 & 0.62 & 31 & 0.72 & 39 & 0.56 & 66 & 0.62 & 102 & 0.60 & 116 & 0.58 \\
Economics, Econometrics and Finance & 21 & 1.64 & 48 & 1.98 & 112 & 2.59 & 184 & 2.65 & 284 & 2.66 & 450 & 2.63 & 541 & 2.70 \\
Energy & 3 & 0.23 & 9 & 0.37 & 19 & 0.44 & 26 & 0.37 & 42 & 0.39 & 110 & 0.64 & 176 & 0.88 \\
Computer Science & 7 & 0.55 & 14 & 0.58 & 58 & 1.34 & 137 & 1.97 & 263 & 2.47 & 518 & 3.03 & 656 & 3.27 \\
Business, Management and Accounting & 3 & 0.23 & 17 & 0.70 & 53 & 1.22 & 148 & 2.13 & 324 & 3.04 & 604 & 3.53 & 708 & 3.53 \\
Health Professions & 3 & 0.23 & 8 & 0.33 & 20 & 0.46 & 38 & 0.55 & 91 & 0.85 & 160 & 0.94 & 180 & 0.90 \\ 
\midrule
in total & 1280 & \multicolumn{1}{l|}{} & 2430 & \multicolumn{1}{l|}{} & 4328 & \multicolumn{1}{l|}{} & 6950 & \multicolumn{1}{l|}{} & 10668 & \multicolumn{1}{l|}{} & 17087 & \multicolumn{1}{l|}{} & 20040 & \multicolumn{1}{l}{} \\
\bottomrule
\end{tabular}
}
\end{table}

\begin{table}[h!]
\centering
\caption{Incorrectly disambiguated periodicals found in MAG. The third column indicates when the data corruption occurred. E.g. ``1990-'' means that the error lasted from the 1950s to the 1980s, and ``2000+'' means it started from the 2000s and last untill the 2010s. }
\label{tab:si:data_corruption}
\resizebox{\linewidth}{!}{%
\begin{tabular}{l|c|c|l} 
\toprule
Periodical Name in MAG & Established & Corrupted & \multicolumn{1}{c}{Mixed up with} \\ 
\midrule
Japanese Journal of Pharmacology              & 1950s & 2010s+ & Journal of Japanese Philosophy \\
Journal of Computers                          & 1950s & 2000s- & Journal de Chimie Physique \\
Journal of Algorithms                         & 1980s & 2010s+ & Jurnal Ilmu Alam dan Lingkungan \\
Journal of Agricultural Engineering Research  & 1960s & 2010s+ & Journal of Advances in Education Research \\
Sozial-und Praventivmedizin                   & 1970s & 2010s+ & Phytothérapie \\
Scientia Forestalis                           & 1950s & 2000s- & Book reviews puslished on Science \\
Interpretation                                & 1970s & 2010s+ & Interpretation \\
Genes                                         & 1990s & 2000s- & Genes \\
Protein Science                               & 1950s & 1990s- & Fortschritte der Physik (Progress of Physics) \\
Hospital Medicine                             & 1990s & 2010s+ & Hospitality  Society \\
Immunotechnology                              & 1990s & 2000s+ & Informacijos mokslai \\
Journal of Ayurveda and Integrative Medicine  & 1990s & 2010s- & The Bulltin of Legal Medicine \\
Versus                                        & 1990s & 2010s- & IEEE Workshop on Visual Surveillance \\
Tradition                                     & 1950s & 1980s+ & Infant Mental Health Journal \\
ACM Transactions on Cyber-Physical Systems    & 2000s & 2010s- & Thermal Conductivity \\
Journal of Biomedical Engineering             & 1970s & 2000s- & Sheng Wu Yi Xue Gong Cheng Xue Za Zhi \\
Antibiotics and Chemotherapy                  & 1950s & 2010s- & Chemotherapy \\
Social Work                                   & 1960s & 2010s+ & Semantic Web \\
Production Journal                            & 1980s & 1990s- & Child Phonology \\
Insight                                       & 1990s & 1990s+ & Insight \\
Sats                                          & 2000s & 2010s+ & SPE Saudi Arabia Section Technical Symposium and Exhibition \\
Leonardo                                      & 1960s & 2010s+ & Innovation and Its Discontents (a book) \\
The Forum                                     & 1990s & 2010s- & Forum \\
Chemical Industry                             & 2000s & 2010s- & Petroleum and Chemical Industry Conference Europe \\
The American review of respiratory disease    & 1950s & 2000s+ & papers missing - only 1 paper in the whole 2000s \\
Chemistry  Industry                           & 1950s & 2010s+ & KEMIJA U INDUSTRIJI (KUI, "Chemistry in Industry"), Chemistry—An Asian Journal \\
Biosilico                                     & 1990s & 2010+  & BIO, and Biodik : Jurnal Ilmiah Pendidikan Biologi \\
Computer Science and Its Applications         & 1990s & 1990s+ & Current Swedish Archaeology \\
Journal of Programming Languages              & 1990s & 2000s+ & Journal of Politics and Law \\
\bottomrule
\end{tabular}
}
\end{table}

\begin{table}[h!]
\centering
\caption{Hyperparameter tuning in the model training. Each model was trained using the same citation trails. The minimum frequency is set to 50 in all settings. $W$ is the context window size, $D$ is the number of embedding dimensions. $\tau$ is the the Kendall rank correlation coefficient between the proportion of papers and $Mean(discipline)/Mean(general)$.}
\label{tab:si:hyperparameter}
\begin{tabular}{c c | c c | c c | c c} 
\toprule
 & & \multicolumn{2}{c|}{\emph{Nature}} & \multicolumn{2}{c|}{\emph{Science}} & \multicolumn{2}{c}{\emph{PNAS}} \\
$D$ & $W$ & $\tau$ & $p$-value & $\tau$ & $p$-value & $\tau$ & $p$-value \\
\midrule
50 & 2 & 0.5270 & $4.85\times10^{-26}$ & 0.5183 & $3.20\times10^{-25}$ & 0.4114 & $4.93\times10^{-16}$\\
50 & 5 & 0.5489 & $4.21\times10^{-28}$ & 0.5251 & $7.24\times10^{-26}$ & 0.4261 & $4.88\times10^{-17}$\\
50 & 10 & 0.5299 & $2.60\times10^{-26}$ & 0.5012 & $1.05\times10^{-23}$ & 0.4017 & $2.56\times10^{-15}$\\
100 & 2 & 0.5475 & $5.59\times10^{-28}$ & 0.5204 & $2.00\times10^{-25}$ & 0.4017 & $2.47\times10^{-15}$\\
100 & 5 & 0.5351 & $8.54\times10^{-27}$ & 0.5098 & $1.81\times10^{-24}$ & 0.3959 & $6.05\times10^{-15}$\\
100 & 10 & 0.5283 & $3.84\times10^{-26}$ & 0.4885 & $1.36\times10^{-22}$ & 0.3939 & $8.73\times10^{-15}$\\
200 & 2 & 0.5408 & $2.43\times10^{-27}$ & 0.5245 & $8.24\times10^{-26}$ & 0.3930 & $9.15\times10^{-15}$\\
200 & 5 & 0.5431 & $1.49\times10^{-27}$ & 0.5059 & $4.03\times10^{-24}$ & 0.3893 & $1.67\times10^{-14}$\\
200 & 10 & 0.5192 & $2.58\times10{-25}$ & 0.4896 & $1.10\times10^{-22}$ & 0.3831 & $4.13\times10^{-14}$\\
300 & 2 & 0.5426 & $1.70\times10^{-27}$ & 0.5178 & $3.42\times10^{-25}$ & 0.3932 & $8.61\times10^{-15}$\\
300 & 5 & 0.5360 & $7.01\times10^{-27}$ & 0.5002 & $1.32\times10^{-23}$ & 0.3869 & $2.35\times10^{-14}$\\
300 & 10 & 0.5386 & $3.98\times10^{-27}$ & 0.4966 & $2.72\times10^{-23}$ & 0.3695 & $2.96\times10^{-13}$ \\
\bottomrule
\end{tabular}
\end{table}

\begin{table}[h!]
\centering
\caption{Top 10 neighbors of \emph{Nature} in different decades.}
\label{tab:si:nature neighbours}
\resizebox{\linewidth}{!}{%
\begin{tabular}{l|l|l|l} 
\toprule
\multicolumn{1}{c|}{1950s} & \multicolumn{1}{c|}{1960s} & \multicolumn{1}{c|}{1970s} & \multicolumn{1}{c}{1980s} \\ 
Journal of Molecular Biology & Science & Science & Science \\
Biochimica et Biophysica Acta & International Geophysics & Advances in Cell Biology & PNAS \\
Naturwissenschaften & Naturwissenschaften & PNAS & Oncogene Research \\
Current Science & Proc. Royal Soc. B & Leukocyte Culture Conference & Oncogene \\
Research & Biochimica et Biophysica Acta & Current Genetics & Progress in Growth Factor Research \\
Bull. Soc. chim. biol. & Journal of Cell Science & Immunological Investigations & The EMBO Journal \\
Methods in Enzymology & Philos. Trans. R. Soc. A & Results and problems in cell differentiation & Cold Spring Harb. Symp. Quant. Biol. \\
New Phytologist & Biochemical Journal & Cell & Current Protocols in Molecular Biology  \\
Trans. R. Soc. South Africa & Indian Journal of Biochemistry & Scottish Journal of Geology & J. Mol. Cell. Immunol. \\
Biochemical Journal & Comprehensive Biochemistry & Cell Biology and Immunology of Leukocyte Function & Cell \\ 
\midrule
\multicolumn{1}{c|}{1990s} & \multicolumn{1}{c|}{2000s} & \multicolumn{1}{c|}{2010s} & \\
Science & Science & Science & \\
PNAS & PNAS & Nature Communication & \\
Current Biology & PLOS Biology & Science Advances & \\
Cell & Reflets De La Physique & PNAS & \\
Evol. Dev. & Harvey Lectures & iScience & \\
J. Mol. Cell. Immunol. & Epigenetics Chromatin & National Science Review & \\
The EMBO Journal & Cold Spring Harb. Perspect. Biol. & Cell & \\
Curr. Opin. Genet. Dev. & Embo Molecular Medicine & Scientific Reports & \\
In Silico Biol. & PLOS ONE Clinical & OMICs & \\
Expert Rev. Mol. Med. & Math. Biol. Bioinform. & Journal of Genomes and Exomes & \\
\bottomrule
\end{tabular}
}
\end{table}
\clearpage

\begin{table}[h!]
\centering
\caption{Top 10 neighbors of \emph{Bulletin of Mathematical Biology} in different decades.}
\label{tab:si:Bull. Math. Biol. neighbours}
\resizebox{\linewidth}{!}{%
\begin{tabular}{l|l|l|l} 
\toprule
\multicolumn{1}{c|}{1950s} & \multicolumn{1}{c|}{1960s} & \multicolumn{1}{c|}{1970s} & \multicolumn{1}{c}{1980s} \\ 
Synthese & J. Theor. Biol. & J. Math. Biol. & Bellman Prize Math. Biosci.  \\
Adv. Biol. Med. Phys. & IEEE Trans. Biomed. Eng. & Bellman Prize Math. Biosci. & J. Math. Biol. \\
Trabajos De Estadistica & Bellman Prize Math. Biosci. & J. Theor. Biol. & J. Theor. Biol. \\
Hereditas & Int. Jt. Conf. Artif. Intell. & Biophys. J. & Math. Model. \\
Ire Trans. Med. Electron. & R. I. Med. J. & Siam J. Appl. Math. & Appl. Math. Lett. \\
Tijdschrift Voor Filosofie & Kybernetika & Biol. Cybern. & Math. Comput. Simul. \\
Psychometrika & BioSystems & Adv. Biol. Med. Phys. & Probab. Eng. Inf. Sci. \\
Arkiv för Matematik & J. Membr. Biol. & Ann. Biomed. Eng. & Math. Med. Biol. \\
Sch. Sci. Math. & J. Biomech. & Ecol. Model. & Phys. D: Nonlinear Phenom. \\
Jpn. J. Physiol. & Inf. Comput. & J. Biol. Phys. & Kybernetes \\ 
\midrule
\multicolumn{1}{c|}{1990s} & \multicolumn{1}{c|}{2000s} & \multicolumn{1}{c|}{2010s} & \\ 
J. Theor. Biol. & J. Math. Biol. & Bellman Prize Math. Biosci. & \\
IEEE IJCNN & J. Theor. Biol. & J. Math. Biol. & \\
J. Biol. Syst. & Math. Med. Biol. & J. Theor. Biol. & \\
Math. Med. Biol. & Math. Biosci. Eng. & Math. Med. Biol. & \\
Artificial Life & Bellman Prize Math. Biosci. & Int. J. Biomath. & \\
Bellman Prize Math. Biosci. & Math. Model. Nat. Phenom. & BioSystems & \\
IEEE Trans. Neural Netw. & Acta Biotheor. & J. Biol. Dyn. & \\
Simulated Evol. Learn. & BioSystems & Math. Biosci. Eng. & \\
J. Math. Biol. & J. Biol. Syst. & Theor. Popul. Biol. & \\
N. Z. Int. Two-Stream Conf. Artif. Neural Netw. Expert Syst. & Theor. Biol. Med. Model. & Eur. Conf. Math. Theor. Biol. & \\
\bottomrule
\end{tabular}
}
\end{table}

\clearpage

\begin{table}[h!]
\centering
\caption{Top 10 neighbors of \emph{AIDS} in different decades.}
\label{tab:si:aids_neighbours}
\resizebox{\linewidth}{!}{%
\begin{tabular}{l | l | l | l } 
\toprule
\multicolumn{1}{c|}{1980s} & \multicolumn{1}{c|}{1990s} & \multicolumn{1}{c|}{2000s} & \multicolumn{1}{c}{2010s} \\ 
\midrule
J. Acquir. Immune Defic. Syndr. & J. Acquir. Immune Defic. Syndr. & J. Acquir. Immune Defic. Syndr. & J. Acquir. Immune Defic. Syndr. \\
AIDS Care & AIDS Res. Hum. Retrovir. & HIV Med. & Lancet HIV \\
Morb. Mortal. Wkly. Rep. & Antivir. Ther. & Antivir. Ther. & Curr. HIV/AIDS Rep. \\
Fam. Pract. & P. R. Health Sci. J. & HIV Clin. Trials & J. Int. AIDS Soc. \\
J. Public Health & AIDS Patient Care STDs & Curr. Opin. HIV AIDS & AIDS Res. Hum. Retrovir. \\
NIPH Ann. & Curr. Opin. Infect. Dis. & AIDS Rev. & HIV Med. \\
Prog. Hematol. & Int. J. STD AIDS & AIDS Res. Hum. Retrovir. & Aids Res. Ther. \\
Del. Med. J. & J. Infect. Dis. & AIDS Care & Curr. Opin. HIV AIDS \\
Boll. Ist. sieroter. milan. & J. Assoc. Nurses AIDS Care & AIDS Patient Care STDs & AIDS Patient Care STDs \\
Pediatr. Infect. Dis. & AIDS Clin. Care & HIV AIDS Rev. & AIDS Rev. \\ 
\bottomrule
\end{tabular}
}
\end{table}

\begin{table}[h!]
\centering
\caption{Periodicals' abbrevations used in Fig.~\ref{fig:si:total_local_semantic_change}.}
\label{tab:si:abbr}
\resizebox{\linewidth}{!}{%
\begin{tabular}{l|l|l|l} 
\toprule
\multicolumn{1}{c|}{Periodical Name} & \multicolumn{1}{c|}{Abbreviation} & \multicolumn{1}{c|}{Periodical Name} & \multicolumn{1}{c}{Abbreviation} \\ 
\midrule
CA: A Cancer Journal for Clinicians & CA: Cancer J. Clin. & Social Networks & Soc. Netw. \\
Quarterly Journal of Economics & QJE & Life sciences in space research & Life Sci Space Res \\
Econometrica & Econometrica & Computational Biology and Chemistry & Comput Biol Chem \\
Psychological Bulletin & Psychol. Bull. & Civil Engineering & C.E.J \\
Chemical Reviews & Chem. Rev. & Journal of Biosciences & J. Biosci. \\
JAMA & JAMA & Cell & Cell \\
Science & Science & Applied Linguistics & Appl. Linguist. \\
Nature & Nature & Journal of Accounting and Economics & J. Account. Econ \\
Proceedings of the National Academy of Sciences of the USA & PNAS & Journal of Accounting and Public Policy & JAPP \\
Physical Review Letters & PRL & Journal of Physics: Condensed Matter & J. Phys. Condens. Matter \\
The New England Journal of Medicine & NEJM & Transport Reviews & Transp. Rev. \\
American Sociological Review & ASR & European Management Journal & EMJ \\
Annals of Mathematics & Ann. Math. & International Journal of Remote Sensing & Int. J. Remote Sens. \\
The Lancet & Lancet & Stem Cells & Stem Cells \\
BMJ & BMJ & Journal of Chemometrics & J. Chemom. \\
Proceedings of The Royal Society B: Biological Sciences & Proc. R. Soc. B & Bioelectromagnetics & Bioelectromagnetics \\
Atmosphere & Atmosphere & neural information processing systems & NeurIPS \\
Language Learning & Lang. Learn. & IEEE Transactions on Medical Imaging & IEEE TMI \\
Automatica & Automatica & The Accounting Review & Account. Rev. \\
Materials Research Bulletin & Mater. Res. Bull. & Human Resource Management Journal & Hum. Resour. Manag. J. \\
Carbon & Carbon & Cancer Cell & Cancer Cell \\
Stanford Law Review & SLR & IEEE Transactions on Applied Superconductivity & IEEE TAS \\
Computing & Computing & Cell Research & Cell Res. \\
Journal of Applied Crystallography & J. Appl. Crystallogr. & the web conference & TheWebConf \\
Ultrasonics & Ultrasonics & knowledge discovery and data mining & KDD \\
IEEE Transactions on Nuclear Science & IEEE Trans Nucl Sci & empirical methods in natural language processing & EMNLP \\
IEEE Transactions on Biomedical Engineering & IEEE. Trans. Biomed. Eng. & Materials & Materials \\
Pattern Recognition & Pattern Recognit. & Sensors & Sensors \\
Physics Letters B & PLB & Complexity & Complexity \\
Journal of Financial and Quantitative Analysis & JFQA & PLOS ONE & PLOS ONE \\
Studies in Second Language Acquisition & Stud. Second Lang. Acquis. & Nature Reviews Immunology & Nat. Rev. Immunol. \\
Linguistic Inquiry & Linguist. Inq. & Nature Materials & Nat. Mater \\
European Journal of Political Research & EJPR & Lancet Oncology & Lancet Oncol. \\
Accounting Organizations and Society & Account. Organ. Soc. & Nature Photonics & Nat. Photonics \\
Clinical Infectious Diseases & Clin. Infect. Dis. & Obesity & Obesity \\
Economic Analysis and Policy & Econ Anal Policy & Nature Chemical Biology & Nat. Chem. Biol. \\
Research Policy & Res. Policy & IEEE Transactions on Industrial Informatics & IEEE TII \\
Gene & Gene & BMC Research Notes & BMC Res. Notes \\ 
Pain & Pain &  &  \\ 
\bottomrule
\end{tabular}
}
\end{table}

\clearpage

\begin{table}[h!]
\centering
\caption{List of the top periodicals with the largest $\Delta d$ in each decade.}
\label{tab:si:top-venue-delta}
\resizebox{0.4\linewidth}{!}{%
\begin{tabular}{lr}
\toprule
Periodical & $\Delta d$ \\
\midrule
\multicolumn{2}{c}{1960s} \\
Bulletin of Environmental Contamination and Toxicology & 0.323 \\
Physical Therapy & 0.320 \\
The Journal of Nuclear Medicine & 0.319 \\
Medical \& Biological Engineering \& Computing & 0.315 \\
Calcified Tissue International & 0.314 \\
Reproduction & 0.302 \\
Clinical Obstetrics and Gynecology & 0.290 \\
European Journal of Nutrition & 0.290 \\
Diabetologia & 0.285 \\
Journal of Catalysis & 0.284 \\
\midrule
\multicolumn{2}{c}{1970s} \\
Synthesis & 0.303 \\
Drug Development and Industrial Pharmacy & 0.296 \\
Pain & 0.293 \\
Journal of Food Science and Technology-mysore & 0.286 \\
Scandinavian Journal of Rheumatology & 0.284 \\
Kidney International & 0.274 \\
Journal of Optics & 0.274 \\
Clinical \& Experimental Allergy & 0.274 \\
International Journal of Pharmaceutics & 0.273 \\
The Journal of Allergy and Clinical Immunology & 0.269 \\
\midrule
\multicolumn{2}{c}{1980s} \\
Sleep & 0.308 \\
Biomaterials & 0.288 \\
AIDS & 0.266 \\
Journal of Controlled Release & 0.265 \\
Applied Organometallic Chemistry & 0.262 \\
Archives of Gerontology and Geriatrics & 0.261 \\
Journal of Automated Methods \& Management in Chemistry & 0.261 \\
Particle \& Particle Systems Characterization & 0.256 \\
Fitoterapia & 0.247 \\
Biomedicine \& Pharmacotherapy & 0.243 \\
\midrule
\multicolumn{2}{c}{1990s} \\
Cell Transplantation & 0.285 \\
Europace & 0.253 \\
Enfermedades Infecciosas Y Microbiologia Clinica & 0.248 \\
Journal of Sleep Research & 0.247 \\
Indicator South Africa & 0.240 \\
International Conference on Telecommunications & 0.237 \\
Nanotechnology & 0.234 \\
Materials Science and Engineering: C & 0.229 \\
Biological Research & 0.224 \\
Experimental and Molecular Medicine & 0.221 \\
\bottomrule
\end{tabular}
}
\end{table}


\begin{thebibliography}{46}%
\makeatletter
\providecommand \@ifxundefined [1]{%
 \@ifx{#1\undefined}
}%
\providecommand \@ifnum [1]{%
 \ifnum #1\expandafter \@firstoftwo
 \else \expandafter \@secondoftwo
 \fi
}%
\providecommand \@ifx [1]{%
 \ifx #1\expandafter \@firstoftwo
 \else \expandafter \@secondoftwo
 \fi
}%
\providecommand \natexlab [1]{#1}%
\providecommand \enquote  [1]{``#1''}%
\providecommand \bibnamefont  [1]{#1}%
\providecommand \bibfnamefont [1]{#1}%
\providecommand \citenamefont [1]{#1}%
\providecommand \href@noop [0]{\@secondoftwo}%
\providecommand \href [0]{\begingroup \@sanitize@url \@href}%
\providecommand \@href[1]{\@@startlink{#1}\@@href}%
\providecommand \@@href[1]{\endgroup#1\@@endlink}%
\providecommand \@sanitize@url [0]{\catcode `\\12\catcode `\$12\catcode
  `\&12\catcode `\#12\catcode `\^12\catcode `\_12\catcode `\%12\relax}%
\providecommand \@@startlink[1]{}%
\providecommand \@@endlink[0]{}%
\providecommand \url  [0]{\begingroup\@sanitize@url \@url }%
\providecommand \@url [1]{\endgroup\@href {#1}{\urlprefix }}%
\providecommand \urlprefix  [0]{URL }%
\providecommand \Eprint [0]{\href }%
\providecommand \doibase [0]{https://doi.org/}%
\providecommand \selectlanguage [0]{\@gobble}%
\providecommand \bibinfo  [0]{\@secondoftwo}%
\providecommand \bibfield  [0]{\@secondoftwo}%
\providecommand \translation [1]{[#1]}%
\providecommand \BibitemOpen [0]{}%
\providecommand \bibitemStop [0]{}%
\providecommand \bibitemNoStop [0]{.\EOS\space}%
\providecommand \EOS [0]{\spacefactor3000\relax}%
\providecommand \BibitemShut  [1]{\csname bibitem#1\endcsname}%
\let\auto@bib@innerbib\@empty
\bibitem [{\citenamefont {Henry}(1665)}]{henry1665epistle}%
  \BibitemOpen
  \bibfield  {author} {\bibinfo {author} {\bibfnamefont {O.}~\bibnamefont
  {Henry}},\ }\bibfield  {title} {\bibinfo {title} {Epistle dedicatory},\
  }\href {https://doi.org/10.1098/rstl.1665.0001} {\bibfield  {journal}
  {\bibinfo  {journal} {Philosophical Transactions of the Royal Society}\
  }\textbf {\bibinfo {volume} {1}},\ \bibinfo {pages} {i} (\bibinfo {year}
  {1665})}\BibitemShut {NoStop}%
\bibitem [{\citenamefont {Baldwin}(2015)}]{baldwin2015making}%
  \BibitemOpen
  \bibfield  {author} {\bibinfo {author} {\bibfnamefont {M.}~\bibnamefont
  {Baldwin}},\ }\href@noop {} {\emph {\bibinfo {title} {Making ``Nature'': The
  History of a Scientific Journal}}}\ (\bibinfo  {publisher} {University of
  Chicago Press},\ \bibinfo {year} {2015})\BibitemShut {NoStop}%
\bibitem [{\citenamefont {Csiszar}(2018)}]{csiszar2018scientific}%
  \BibitemOpen
  \bibfield  {author} {\bibinfo {author} {\bibfnamefont {A.}~\bibnamefont
  {Csiszar}},\ }\href@noop {} {\emph {\bibinfo {title} {The Scientific Journal:
  Authorship and the Politics of Knowledge in the Nineteenth Century}}}\
  (\bibinfo  {publisher} {University of Chicago Press},\ \bibinfo {year}
  {2018})\BibitemShut {NoStop}%
\bibitem [{\citenamefont {Uzzi}\ \emph {et~al.}(2013)\citenamefont {Uzzi},
  \citenamefont {Mukherjee}, \citenamefont {Stringer},\ and\ \citenamefont
  {Jones}}]{uzzi2013atypical}%
  \BibitemOpen
  \bibfield  {author} {\bibinfo {author} {\bibfnamefont {B.}~\bibnamefont
  {Uzzi}}, \bibinfo {author} {\bibfnamefont {S.}~\bibnamefont {Mukherjee}},
  \bibinfo {author} {\bibfnamefont {M.}~\bibnamefont {Stringer}},\ and\
  \bibinfo {author} {\bibfnamefont {B.}~\bibnamefont {Jones}},\ }\bibfield
  {title} {\bibinfo {title} {Atypical combinations and scientific impact},\
  }\href {https://doi.org/10.1126/science.1240474} {\bibfield  {journal}
  {\bibinfo  {journal} {Science}\ }\textbf {\bibinfo {volume} {342}},\ \bibinfo
  {pages} {468} (\bibinfo {year} {2013})}\BibitemShut {NoStop}%
\bibitem [{\citenamefont {Wang}\ \emph {et~al.}(2017)\citenamefont {Wang},
  \citenamefont {Veugelers},\ and\ \citenamefont {Stephan}}]{wang2017bias}%
  \BibitemOpen
  \bibfield  {author} {\bibinfo {author} {\bibfnamefont {J.}~\bibnamefont
  {Wang}}, \bibinfo {author} {\bibfnamefont {R.}~\bibnamefont {Veugelers}},\
  and\ \bibinfo {author} {\bibfnamefont {P.}~\bibnamefont {Stephan}},\
  }\bibfield  {title} {\bibinfo {title} {Bias against novelty in science: A
  cautionary tale for users of bibliometric indicators},\ }\href
  {https://doi.org/10.1016/j.respol.2017.06.006} {\bibfield  {journal}
  {\bibinfo  {journal} {Research Policy}\ }\textbf {\bibinfo {volume} {46}},\
  \bibinfo {pages} {1416} (\bibinfo {year} {2017})}\BibitemShut {NoStop}%
\bibitem [{\citenamefont {Narin}\ \emph {et~al.}(1976)\citenamefont {Narin},
  \citenamefont {Pinski},\ and\ \citenamefont {Gee}}]{narin1976structure}%
  \BibitemOpen
  \bibfield  {author} {\bibinfo {author} {\bibfnamefont {F.}~\bibnamefont
  {Narin}}, \bibinfo {author} {\bibfnamefont {G.}~\bibnamefont {Pinski}},\ and\
  \bibinfo {author} {\bibfnamefont {H.~H.}\ \bibnamefont {Gee}},\ }\bibfield
  {title} {\bibinfo {title} {Structure of the biomedical literature},\ }\href
  {https://doi.org/10.1002/asi.4630270104} {\bibfield  {journal} {\bibinfo
  {journal} {Journal of the American Society for Information Science}\ }\textbf
  {\bibinfo {volume} {27}},\ \bibinfo {pages} {25} (\bibinfo {year}
  {1976})}\BibitemShut {NoStop}%
\bibitem [{\citenamefont {N{\'u}{\~n}ez}\ \emph {et~al.}(2019)\citenamefont
  {N{\'u}{\~n}ez}, \citenamefont {Allen}, \citenamefont {Gao}, \citenamefont
  {Miller~Rigoli}, \citenamefont {Relaford-Doyle},\ and\ \citenamefont
  {Semenuks}}]{nunez2019happened}%
  \BibitemOpen
  \bibfield  {author} {\bibinfo {author} {\bibfnamefont {R.}~\bibnamefont
  {N{\'u}{\~n}ez}}, \bibinfo {author} {\bibfnamefont {M.}~\bibnamefont
  {Allen}}, \bibinfo {author} {\bibfnamefont {R.}~\bibnamefont {Gao}}, \bibinfo
  {author} {\bibfnamefont {C.}~\bibnamefont {Miller~Rigoli}}, \bibinfo {author}
  {\bibfnamefont {J.}~\bibnamefont {Relaford-Doyle}},\ and\ \bibinfo {author}
  {\bibfnamefont {A.}~\bibnamefont {Semenuks}},\ }\bibfield  {title} {\bibinfo
  {title} {What happened to cognitive science?},\ }\href
  {https://doi.org/10.1038/s41562-019-0626-2} {\bibfield  {journal} {\bibinfo
  {journal} {Nature Human Behaviour}\ }\textbf {\bibinfo {volume} {3}},\
  \bibinfo {pages} {782} (\bibinfo {year} {2019})}\BibitemShut {NoStop}%
\bibitem [{\citenamefont {Line}(1970)}]{line1970half}%
  \BibitemOpen
  \bibfield  {author} {\bibinfo {author} {\bibfnamefont {M.~B.}\ \bibnamefont
  {Line}},\ }\bibfield  {title} {\bibinfo {title} {The ‘half-life’ of
  periodical literature: Apparent and real obsolescence},\ }\href
  {https://doi.org/10.1108/eb026486} {\bibfield  {journal} {\bibinfo  {journal}
  {Journal of Documentation}\ }\textbf {\bibinfo {volume} {26}},\ \bibinfo
  {pages} {46} (\bibinfo {year} {1970})}\BibitemShut {NoStop}%
\bibitem [{\citenamefont {Calcagno}\ \emph {et~al.}(2012)\citenamefont
  {Calcagno}, \citenamefont {Demoinet}, \citenamefont {Gollner}, \citenamefont
  {Guidi}, \citenamefont {Ruths},\ and\ \citenamefont
  {de~Mazancourt}}]{calcagno2012flows}%
  \BibitemOpen
  \bibfield  {author} {\bibinfo {author} {\bibfnamefont {V.}~\bibnamefont
  {Calcagno}}, \bibinfo {author} {\bibfnamefont {E.}~\bibnamefont {Demoinet}},
  \bibinfo {author} {\bibfnamefont {K.}~\bibnamefont {Gollner}}, \bibinfo
  {author} {\bibfnamefont {L.}~\bibnamefont {Guidi}}, \bibinfo {author}
  {\bibfnamefont {D.}~\bibnamefont {Ruths}},\ and\ \bibinfo {author}
  {\bibfnamefont {C.}~\bibnamefont {de~Mazancourt}},\ }\bibfield  {title}
  {\bibinfo {title} {Flows of research manuscripts among scientific journals
  reveal hidden submission patterns},\ }\href
  {https://doi.org/10.1126/science.1227833} {\bibfield  {journal} {\bibinfo
  {journal} {Science}\ }\textbf {\bibinfo {volume} {338}},\ \bibinfo {pages}
  {1065} (\bibinfo {year} {2012})}\BibitemShut {NoStop}%
\bibitem [{\citenamefont {Leydesdorff}(1987)}]{leydesdorff1987various}%
  \BibitemOpen
  \bibfield  {author} {\bibinfo {author} {\bibfnamefont {L.}~\bibnamefont
  {Leydesdorff}},\ }\bibfield  {title} {\bibinfo {title} {Various methods for
  the mapping of science},\ }\href@noop {} {\bibfield  {journal} {\bibinfo
  {journal} {Scientometrics}\ }\textbf {\bibinfo {volume} {11}},\ \bibinfo
  {pages} {295} (\bibinfo {year} {1987})}\BibitemShut {NoStop}%
\bibitem [{\citenamefont {Small}(1999)}]{small1999visualizing}%
  \BibitemOpen
  \bibfield  {author} {\bibinfo {author} {\bibfnamefont {H.}~\bibnamefont
  {Small}},\ }\bibfield  {title} {\bibinfo {title} {Visualizing science by
  citation mapping},\ }\href
  {https://doi.org/10.1002/(SICI)1097-4571(1999)50:9<799::AID-ASI9>3.0.CO;2-G}
  {\bibfield  {journal} {\bibinfo  {journal} {Journal of the American Society
  for Information Science}\ }\textbf {\bibinfo {volume} {50}},\ \bibinfo
  {pages} {799} (\bibinfo {year} {1999})}\BibitemShut {NoStop}%
\bibitem [{\citenamefont {Shiffrin}\ and\ \citenamefont
  {B\"{o}rner}(2004)}]{shiffrin2004mapping}%
  \BibitemOpen
  \bibfield  {author} {\bibinfo {author} {\bibfnamefont {R.~M.}\ \bibnamefont
  {Shiffrin}}\ and\ \bibinfo {author} {\bibfnamefont {K.}~\bibnamefont
  {B\"{o}rner}},\ }\bibfield  {title} {\bibinfo {title} {Mapping knowledge
  domains},\ }\href {https://doi.org/10.1073/pnas.0307852100} {\bibfield
  {journal} {\bibinfo  {journal} {Proceedings of the National Academy of
  Sciences}\ }\textbf {\bibinfo {volume} {101}},\ \bibinfo {pages} {5183}
  (\bibinfo {year} {2004})}\BibitemShut {NoStop}%
\bibitem [{\citenamefont {Boyack}\ \emph {et~al.}(2005)\citenamefont {Boyack},
  \citenamefont {Klavans},\ and\ \citenamefont
  {B{\"o}rner}}]{boyack2005mapping}%
  \BibitemOpen
  \bibfield  {author} {\bibinfo {author} {\bibfnamefont {K.~W.}\ \bibnamefont
  {Boyack}}, \bibinfo {author} {\bibfnamefont {R.}~\bibnamefont {Klavans}},\
  and\ \bibinfo {author} {\bibfnamefont {K.}~\bibnamefont {B{\"o}rner}},\
  }\bibfield  {title} {\bibinfo {title} {Mapping the backbone of science},\
  }\href {https://doi.org/10.1007/s11192-005-0255-6} {\bibfield  {journal}
  {\bibinfo  {journal} {Scientometrics}\ }\textbf {\bibinfo {volume} {64}},\
  \bibinfo {pages} {351} (\bibinfo {year} {2005})}\BibitemShut {NoStop}%
\bibitem [{\citenamefont {Rosvall}\ and\ \citenamefont
  {Bergstrom}(2008)}]{rosvall2008maps}%
  \BibitemOpen
  \bibfield  {author} {\bibinfo {author} {\bibfnamefont {M.}~\bibnamefont
  {Rosvall}}\ and\ \bibinfo {author} {\bibfnamefont {C.~T.}\ \bibnamefont
  {Bergstrom}},\ }\bibfield  {title} {\bibinfo {title} {Maps of random walks on
  complex networks reveal community structure},\ }\href
  {https://doi.org/10.1073/pnas.0706851105} {\bibfield  {journal} {\bibinfo
  {journal} {Proceedings of the National Academy of Sciences}\ }\textbf
  {\bibinfo {volume} {105}},\ \bibinfo {pages} {1118} (\bibinfo {year}
  {2008})}\BibitemShut {NoStop}%
\bibitem [{\citenamefont {B{\"o}rner}\ \emph {et~al.}(2012)\citenamefont
  {B{\"o}rner}, \citenamefont {Klavans}, \citenamefont {Patek}, \citenamefont
  {Zoss}, \citenamefont {Biberstine}, \citenamefont {Light}, \citenamefont
  {Larivi{\`e}re},\ and\ \citenamefont {Boyack}}]{borner2012design}%
  \BibitemOpen
  \bibfield  {author} {\bibinfo {author} {\bibfnamefont {K.}~\bibnamefont
  {B{\"o}rner}}, \bibinfo {author} {\bibfnamefont {R.}~\bibnamefont {Klavans}},
  \bibinfo {author} {\bibfnamefont {M.}~\bibnamefont {Patek}}, \bibinfo
  {author} {\bibfnamefont {A.~M.}\ \bibnamefont {Zoss}}, \bibinfo {author}
  {\bibfnamefont {J.~R.}\ \bibnamefont {Biberstine}}, \bibinfo {author}
  {\bibfnamefont {R.~P.}\ \bibnamefont {Light}}, \bibinfo {author}
  {\bibfnamefont {V.}~\bibnamefont {Larivi{\`e}re}},\ and\ \bibinfo {author}
  {\bibfnamefont {K.~W.}\ \bibnamefont {Boyack}},\ }\bibfield  {title}
  {\bibinfo {title} {Design and update of a classification system: The {UCSD}
  map of science},\ }\href {https://doi.org/10.1371/journal.pone.0039464}
  {\bibfield  {journal} {\bibinfo  {journal} {PLOS ONE}\ }\textbf {\bibinfo
  {volume} {7}},\ \bibinfo {pages} {e39464} (\bibinfo {year}
  {2012})}\BibitemShut {NoStop}%
\bibitem [{\citenamefont {Bollen}\ \emph {et~al.}(2009)\citenamefont {Bollen},
  \citenamefont {de~Sompel}, \citenamefont {Hagberg}, \citenamefont
  {Bettencourt}, \citenamefont {Chute}, \citenamefont {Rodriguez},\ and\
  \citenamefont {Balakireva}}]{bollen2009clickstream}%
  \BibitemOpen
  \bibfield  {author} {\bibinfo {author} {\bibfnamefont {J.}~\bibnamefont
  {Bollen}}, \bibinfo {author} {\bibfnamefont {H.~V.}\ \bibnamefont
  {de~Sompel}}, \bibinfo {author} {\bibfnamefont {A.}~\bibnamefont {Hagberg}},
  \bibinfo {author} {\bibfnamefont {L.}~\bibnamefont {Bettencourt}}, \bibinfo
  {author} {\bibfnamefont {R.}~\bibnamefont {Chute}}, \bibinfo {author}
  {\bibfnamefont {M.~A.}\ \bibnamefont {Rodriguez}},\ and\ \bibinfo {author}
  {\bibfnamefont {L.}~\bibnamefont {Balakireva}},\ }\bibfield  {title}
  {\bibinfo {title} {Clickstream data yields high-resolution maps of science},\
  }\href {https://doi.org/10.1371/journal.pone.0004803} {\bibfield  {journal}
  {\bibinfo  {journal} {PLOS ONE}\ }\textbf {\bibinfo {volume} {4}},\ \bibinfo
  {pages} {e4803} (\bibinfo {year} {2009})}\BibitemShut {NoStop}%
\bibitem [{\citenamefont {Peng}\ \emph {et~al.}(2021)\citenamefont {Peng},
  \citenamefont {Ke}, \citenamefont {Budak}, \citenamefont {Romero},\ and\
  \citenamefont {Ahn}}]{peng2021neural}%
  \BibitemOpen
  \bibfield  {author} {\bibinfo {author} {\bibfnamefont {H.}~\bibnamefont
  {Peng}}, \bibinfo {author} {\bibfnamefont {Q.}~\bibnamefont {Ke}}, \bibinfo
  {author} {\bibfnamefont {C.}~\bibnamefont {Budak}}, \bibinfo {author}
  {\bibfnamefont {D.~M.}\ \bibnamefont {Romero}},\ and\ \bibinfo {author}
  {\bibfnamefont {Y.-Y.}\ \bibnamefont {Ahn}},\ }\bibfield  {title} {\bibinfo
  {title} {Neural embeddings of scholarly periodicals reveal complex
  disciplinary organizations},\ }\href {https://doi.org/10.1126/sciadv.abb9004}
  {\bibfield  {journal} {\bibinfo  {journal} {Science Advances}\ }\textbf
  {\bibinfo {volume} {7}},\ \bibinfo {pages} {eabb9004} (\bibinfo {year}
  {2021})}\BibitemShut {NoStop}%
\bibitem [{\citenamefont {Gates}\ \emph
  {et~al.}(2019{\natexlab{a}})\citenamefont {Gates}, \citenamefont {Ke},
  \citenamefont {Varol},\ and\ \citenamefont {Barab{\'a}si}}]{gates2019nature}%
  \BibitemOpen
  \bibfield  {author} {\bibinfo {author} {\bibfnamefont {A.~J.}\ \bibnamefont
  {Gates}}, \bibinfo {author} {\bibfnamefont {Q.}~\bibnamefont {Ke}}, \bibinfo
  {author} {\bibfnamefont {O.}~\bibnamefont {Varol}},\ and\ \bibinfo {author}
  {\bibfnamefont {A.-L.}\ \bibnamefont {Barab{\'a}si}},\ }\bibfield  {title}
  {\bibinfo {title} {Nature’s reach: narrow work has broad impact},\
  }\href@noop {} {\bibfield  {journal} {\bibinfo  {journal} {Nature}\ }\textbf
  {\bibinfo {volume} {575}},\ \bibinfo {pages} {32} (\bibinfo {year}
  {2019}{\natexlab{a}})}\BibitemShut {NoStop}%
\bibitem [{\citenamefont {McGillivray}\ \emph {et~al.}(2022)\citenamefont
  {McGillivray}, \citenamefont {Jenset}, \citenamefont {Salama},\ and\
  \citenamefont {Schut}}]{mcgillivray2022investigating}%
  \BibitemOpen
  \bibfield  {author} {\bibinfo {author} {\bibfnamefont {B.}~\bibnamefont
  {McGillivray}}, \bibinfo {author} {\bibfnamefont {G.~B.}\ \bibnamefont
  {Jenset}}, \bibinfo {author} {\bibfnamefont {K.}~\bibnamefont {Salama}},\
  and\ \bibinfo {author} {\bibfnamefont {D.}~\bibnamefont {Schut}},\ }\bibfield
   {title} {\bibinfo {title} {Investigating patterns of change, stability, and
  interaction among scientific disciplines using embeddings},\ }\href
  {https://doi.org/10.1057/s41599-022-01267-5} {\bibfield  {journal} {\bibinfo
  {journal} {Humanities and Social Sciences Communications}\ }\textbf {\bibinfo
  {volume} {9}},\ \bibinfo {pages} {285} (\bibinfo {year} {2022})}\BibitemShut
  {NoStop}%
\bibitem [{\citenamefont {Mikolov}\ \emph
  {et~al.}(2013{\natexlab{a}})\citenamefont {Mikolov}, \citenamefont {Chen},
  \citenamefont {Corrado},\ and\ \citenamefont {Dean}}]{mikolov2013efficient}%
  \BibitemOpen
  \bibfield  {author} {\bibinfo {author} {\bibfnamefont {T.}~\bibnamefont
  {Mikolov}}, \bibinfo {author} {\bibfnamefont {K.}~\bibnamefont {Chen}},
  \bibinfo {author} {\bibfnamefont {G.}~\bibnamefont {Corrado}},\ and\ \bibinfo
  {author} {\bibfnamefont {J.}~\bibnamefont {Dean}},\ }\bibfield  {title}
  {\bibinfo {title} {Efficient estimation of word representations in vector
  space},\ }\href@noop {} {\bibfield  {journal} {\bibinfo  {journal}
  {arXiv:1301.3781}\ } (\bibinfo {year} {2013}{\natexlab{a}})}\BibitemShut
  {NoStop}%
\bibitem [{\citenamefont {Shoemark}\ \emph {et~al.}(2019)\citenamefont
  {Shoemark}, \citenamefont {Liza}, \citenamefont {Nguyen}, \citenamefont
  {Hale},\ and\ \citenamefont {McGillivray}}]{shoemark2019room}%
  \BibitemOpen
  \bibfield  {author} {\bibinfo {author} {\bibfnamefont {P.}~\bibnamefont
  {Shoemark}}, \bibinfo {author} {\bibfnamefont {F.~F.}\ \bibnamefont {Liza}},
  \bibinfo {author} {\bibfnamefont {D.}~\bibnamefont {Nguyen}}, \bibinfo
  {author} {\bibfnamefont {S.}~\bibnamefont {Hale}},\ and\ \bibinfo {author}
  {\bibfnamefont {B.}~\bibnamefont {McGillivray}},\ }\bibfield  {title}
  {\bibinfo {title} {Room to {G}lo: A systematic comparison of semantic change
  detection approaches with word embeddings},\ }in\ \href
  {https://doi.org/10.18653/v1/D19-1007} {\emph {\bibinfo {booktitle}
  {Proceedings of the 2019 Conference on Empirical Methods in Natural Language
  Processing and the 9th International Joint Conference on Natural Language
  Processing (EMNLP-IJCNLP)}}}\ (\bibinfo {year} {2019})\ pp.\ \bibinfo {pages}
  {66--76}\BibitemShut {NoStop}%
\bibitem [{\citenamefont {Marr}\ and\ \citenamefont
  {Hildreth}(1980)}]{marr1980edgedetection}%
  \BibitemOpen
  \bibfield  {author} {\bibinfo {author} {\bibfnamefont {D.}~\bibnamefont
  {Marr}}\ and\ \bibinfo {author} {\bibfnamefont {E.}~\bibnamefont
  {Hildreth}},\ }\bibfield  {title} {\bibinfo {title} {Theory of edge
  detection},\ }\href {https://doi.org/10.1098/rspb.1980.0020} {\bibfield
  {journal} {\bibinfo  {journal} {Proceedings of the Royal Society of London.
  Series B, Containing papers of a Biological character. Royal Society (Great
  Britain)}\ }\textbf {\bibinfo {volume} {207}},\ \bibinfo {pages} {187}
  (\bibinfo {year} {1980})}\BibitemShut {NoStop}%
\bibitem [{\citenamefont {Longuet-Higgins}\ and\ \citenamefont
  {Prazdny}(1980)}]{LonguetHiggins1980movingretinal}%
  \BibitemOpen
  \bibfield  {author} {\bibinfo {author} {\bibfnamefont {H.~C.}\ \bibnamefont
  {Longuet-Higgins}}\ and\ \bibinfo {author} {\bibfnamefont {K.}~\bibnamefont
  {Prazdny}},\ }\bibfield  {title} {\bibinfo {title} {The interpretation of a
  moving retinal image},\ }\href@noop {} {\bibfield  {journal} {\bibinfo
  {journal} {Proceedings of the Royal Society of London. Series B. Biological
  Sciences}\ }\textbf {\bibinfo {volume} {208}},\ \bibinfo {pages} {385}
  (\bibinfo {year} {1980})}\BibitemShut {NoStop}%
\bibitem [{\citenamefont {Srinivasan}\ \emph {et~al.}(1982)\citenamefont
  {Srinivasan}, \citenamefont {Laughlin},\ and\ \citenamefont
  {Dubs}}]{Srinivasan1982predictivecoding}%
  \BibitemOpen
  \bibfield  {author} {\bibinfo {author} {\bibfnamefont {M.~V.}\ \bibnamefont
  {Srinivasan}}, \bibinfo {author} {\bibfnamefont {S.~B.}\ \bibnamefont
  {Laughlin}},\ and\ \bibinfo {author} {\bibfnamefont {A.}~\bibnamefont
  {Dubs}},\ }\bibfield  {title} {\bibinfo {title} {Predictive coding: a fresh
  view of inhibition in the retina},\ }\href@noop {} {\bibfield  {journal}
  {\bibinfo  {journal} {Proceedings of the Royal Society of London. Series B.
  Biological Sciences}\ }\textbf {\bibinfo {volume} {216}},\ \bibinfo {pages}
  {427} (\bibinfo {year} {1982})}\BibitemShut {NoStop}%
\bibitem [{\citenamefont {Perrett}\ \emph {et~al.}(1985)\citenamefont
  {Perrett}, \citenamefont {Smith}, \citenamefont {Potter}, \citenamefont
  {Mistlin}, \citenamefont {Head}, \citenamefont {Milner},\ and\ \citenamefont
  {Jeeves}}]{Perrett1985visualcells}%
  \BibitemOpen
  \bibfield  {author} {\bibinfo {author} {\bibfnamefont {D.~I.}\ \bibnamefont
  {Perrett}}, \bibinfo {author} {\bibfnamefont {P.~A.~J.}\ \bibnamefont
  {Smith}}, \bibinfo {author} {\bibfnamefont {D.~D.}\ \bibnamefont {Potter}},
  \bibinfo {author} {\bibfnamefont {A.~J.}\ \bibnamefont {Mistlin}}, \bibinfo
  {author} {\bibfnamefont {A.~S.}\ \bibnamefont {Head}}, \bibinfo {author}
  {\bibfnamefont {A.~D.}\ \bibnamefont {Milner}},\ and\ \bibinfo {author}
  {\bibfnamefont {M.~A.}\ \bibnamefont {Jeeves}},\ }\bibfield  {title}
  {\bibinfo {title} {Visual cells in the temporal cortex sensitive to face view
  and gaze direction},\ }\href@noop {} {\bibfield  {journal} {\bibinfo
  {journal} {Proceedings of the Royal Society of London. Series B. Biological
  Sciences}\ }\textbf {\bibinfo {volume} {223}},\ \bibinfo {pages} {293}
  (\bibinfo {year} {1985})}\BibitemShut {NoStop}%
\bibitem [{\citenamefont {Morrone}\ and\ \citenamefont
  {Burr}(1988)}]{Morrone1988featuredetection}%
  \BibitemOpen
  \bibfield  {author} {\bibinfo {author} {\bibfnamefont {M.~C.}\ \bibnamefont
  {Morrone}}\ and\ \bibinfo {author} {\bibfnamefont {D.~C.}\ \bibnamefont
  {Burr}},\ }\bibfield  {title} {\bibinfo {title} {Feature detection in human
  vision: a phase-dependent energy model},\ }\href@noop {} {\bibfield
  {journal} {\bibinfo  {journal} {Proceedings of the Royal Society of London.
  Series B. Biological Sciences}\ }\textbf {\bibinfo {volume} {235}},\ \bibinfo
  {pages} {221} (\bibinfo {year} {1988})}\BibitemShut {NoStop}%
\bibitem [{\citenamefont {Marr}\ and\ \citenamefont
  {Ullman}(1981)}]{Marr1981Directionalselectivity}%
  \BibitemOpen
  \bibfield  {author} {\bibinfo {author} {\bibfnamefont {D.~C.}\ \bibnamefont
  {Marr}}\ and\ \bibinfo {author} {\bibfnamefont {S.}~\bibnamefont {Ullman}},\
  }\bibfield  {title} {\bibinfo {title} {Directional selectivity and its use in
  early visual processing},\ }\href@noop {} {\bibfield  {journal} {\bibinfo
  {journal} {Proceedings of the Royal Society of London. Series B. Biological
  Sciences}\ }\textbf {\bibinfo {volume} {211}},\ \bibinfo {pages} {151}
  (\bibinfo {year} {1981})}\BibitemShut {NoStop}%
\bibitem [{\citenamefont {Gregory}(1980)}]{gregory1980perceptions}%
  \BibitemOpen
  \bibfield  {author} {\bibinfo {author} {\bibfnamefont {R.~L.}\ \bibnamefont
  {Gregory}},\ }\bibfield  {title} {\bibinfo {title} {Perceptions as
  hypotheses},\ }\href@noop {} {\bibfield  {journal} {\bibinfo  {journal}
  {Philosophical Transactions of the Royal Society of London. B, Biological
  Sciences}\ }\textbf {\bibinfo {volume} {290}},\ \bibinfo {pages} {181}
  (\bibinfo {year} {1980})}\BibitemShut {NoStop}%
\bibitem [{\citenamefont {Posner}\ \emph {et~al.}(1982)\citenamefont {Posner},
  \citenamefont {Cohen},\ and\ \citenamefont {Rafal}}]{posner1982neural}%
  \BibitemOpen
  \bibfield  {author} {\bibinfo {author} {\bibfnamefont {M.~I.}\ \bibnamefont
  {Posner}}, \bibinfo {author} {\bibfnamefont {Y.}~\bibnamefont {Cohen}},\ and\
  \bibinfo {author} {\bibfnamefont {R.~D.}\ \bibnamefont {Rafal}},\ }\bibfield
  {title} {\bibinfo {title} {Neural systems control of spatial orienting},\
  }\href@noop {} {\bibfield  {journal} {\bibinfo  {journal} {Philosophical
  Transactions of the Royal Society of London. B, Biological Sciences}\
  }\textbf {\bibinfo {volume} {298}},\ \bibinfo {pages} {187} (\bibinfo {year}
  {1982})}\BibitemShut {NoStop}%
\bibitem [{\citenamefont {Lee}(1980)}]{lee1980optic}%
  \BibitemOpen
  \bibfield  {author} {\bibinfo {author} {\bibfnamefont {D.~N.}\ \bibnamefont
  {Lee}},\ }\bibfield  {title} {\bibinfo {title} {The optic flow field: The
  foundation of vision},\ }\href@noop {} {\bibfield  {journal} {\bibinfo
  {journal} {Philosophical Transactions of the Royal Society of London. B,
  Biological Sciences}\ }\textbf {\bibinfo {volume} {290}},\ \bibinfo {pages}
  {169} (\bibinfo {year} {1980})}\BibitemShut {NoStop}%
\bibitem [{\citenamefont {Balaban}\ and\ \citenamefont
  {Klein}(2006)}]{balaban2006chemistry}%
  \BibitemOpen
  \bibfield  {author} {\bibinfo {author} {\bibfnamefont {A.~T.}\ \bibnamefont
  {Balaban}}\ and\ \bibinfo {author} {\bibfnamefont {D.~J.}\ \bibnamefont
  {Klein}},\ }\bibfield  {title} {\bibinfo {title} {Is chemistry `the central
  science'? how are different sciences related? co-citations, reductionism,
  emergence, and posets},\ }\href {https://doi.org/10.1007/s11192-006-0173-2}
  {\bibfield  {journal} {\bibinfo  {journal} {Scientometrics}\ }\textbf
  {\bibinfo {volume} {69}},\ \bibinfo {pages} {615} (\bibinfo {year}
  {2006})}\BibitemShut {NoStop}%
\bibitem [{\citenamefont {Szell}\ \emph {et~al.}(2018)\citenamefont {Szell},
  \citenamefont {Ma},\ and\ \citenamefont {Sinatra}}]{szell2018nobel}%
  \BibitemOpen
  \bibfield  {author} {\bibinfo {author} {\bibfnamefont {M.}~\bibnamefont
  {Szell}}, \bibinfo {author} {\bibfnamefont {Y.}~\bibnamefont {Ma}},\ and\
  \bibinfo {author} {\bibfnamefont {R.}~\bibnamefont {Sinatra}},\ }\bibfield
  {title} {\bibinfo {title} {A nobel opportunity for interdisciplinarity},\
  }\href {https://doi.org/10.1038/s41567-018-0314-6} {\bibfield  {journal}
  {\bibinfo  {journal} {Nature Physics}\ }\textbf {\bibinfo {volume} {14}},\
  \bibinfo {pages} {1075} (\bibinfo {year} {2018})}\BibitemShut {NoStop}%
\bibitem [{\citenamefont {Gates}\ \emph
  {et~al.}(2019{\natexlab{b}})\citenamefont {Gates}, \citenamefont {Wood},
  \citenamefont {Hetrick},\ and\ \citenamefont {Ahn}}]{gates2019element}%
  \BibitemOpen
  \bibfield  {author} {\bibinfo {author} {\bibfnamefont {A.~J.}\ \bibnamefont
  {Gates}}, \bibinfo {author} {\bibfnamefont {I.~B.}\ \bibnamefont {Wood}},
  \bibinfo {author} {\bibfnamefont {W.~P.}\ \bibnamefont {Hetrick}},\ and\
  \bibinfo {author} {\bibfnamefont {Y.-Y.}\ \bibnamefont {Ahn}},\ }\bibfield
  {title} {\bibinfo {title} {Element-centric clustering comparison unifies
  overlaps and hierarchy},\ }\href@noop {} {\bibfield  {journal} {\bibinfo
  {journal} {Sci. Rep.}\ }\textbf {\bibinfo {volume} {9}},\ \bibinfo {pages}
  {8574} (\bibinfo {year} {2019}{\natexlab{b}})}\BibitemShut {NoStop}%
\bibitem [{\citenamefont {Shepard}(1968)}]{shepard1968two}%
  \BibitemOpen
  \bibfield  {author} {\bibinfo {author} {\bibfnamefont {D.}~\bibnamefont
  {Shepard}},\ }\bibfield  {title} {\bibinfo {title} {A two-dimensional
  interpolation function for irregularly-spaced data},\ }in\ \href
  {https://doi.org/10.1145/800186.810616} {\emph {\bibinfo {booktitle}
  {Proceedings of the 1968 23rd ACM National Conference}}}\ (\bibinfo {year}
  {1968})\ pp.\ \bibinfo {pages} {517--524}\BibitemShut {NoStop}%
\bibitem [{\citenamefont {Hill}(1956)}]{hill1956biophysics}%
  \BibitemOpen
  \bibfield  {author} {\bibinfo {author} {\bibfnamefont {A.~V.}\ \bibnamefont
  {Hill}},\ }\bibfield  {title} {\bibinfo {title} {Why biophysics?},\
  }\href@noop {} {\bibfield  {journal} {\bibinfo  {journal} {Science}\ }\textbf
  {\bibinfo {volume} {124}},\ \bibinfo {pages} {1233} (\bibinfo {year}
  {1956})}\BibitemShut {NoStop}%
\bibitem [{\citenamefont {Woese}(2004)}]{woese2004new}%
  \BibitemOpen
  \bibfield  {author} {\bibinfo {author} {\bibfnamefont {C.~R.}\ \bibnamefont
  {Woese}},\ }\bibfield  {title} {\bibinfo {title} {A new biology for a new
  century},\ }\href {https://doi.org/10.1128/MMBR.68.2.173-186.2004} {\bibfield
   {journal} {\bibinfo  {journal} {Microbiology and Molecular Biology Reviews}\
  }\textbf {\bibinfo {volume} {68}},\ \bibinfo {pages} {173} (\bibinfo {year}
  {2004})}\BibitemShut {NoStop}%
\bibitem [{\citenamefont {L\"{o}wy}(2011)}]{lowy2011historiography}%
  \BibitemOpen
  \bibfield  {author} {\bibinfo {author} {\bibfnamefont {I.}~\bibnamefont
  {L\"{o}wy}},\ }\bibfield  {title} {\bibinfo {title} {Historiography of
  biomedicine: “bio,” “medicine,” and in between},\ }\href
  {https://doi.org/10.1086/658661} {\bibfield  {journal} {\bibinfo  {journal}
  {Isis}\ }\textbf {\bibinfo {volume} {102}},\ \bibinfo {pages} {116} (\bibinfo
  {year} {2011})}\BibitemShut {NoStop}%
\bibitem [{\citenamefont {Monroe}\ \emph {et~al.}(2017)\citenamefont {Monroe},
  \citenamefont {Colaresi},\ and\ \citenamefont {Quinn}}]{monroe2017fightin}%
  \BibitemOpen
  \bibfield  {author} {\bibinfo {author} {\bibfnamefont {B.~L.}\ \bibnamefont
  {Monroe}}, \bibinfo {author} {\bibfnamefont {M.~P.}\ \bibnamefont
  {Colaresi}},\ and\ \bibinfo {author} {\bibfnamefont {K.~M.}\ \bibnamefont
  {Quinn}},\ }\bibfield  {title} {\bibinfo {title} {Fightin' words: Lexical
  feature selection and evaluation for identifying the content of political
  conflict},\ }\href {https://doi.org/10.1093/pan/mpn018} {\bibfield  {journal}
  {\bibinfo  {journal} {Political Analysis}\ }\textbf {\bibinfo {volume}
  {16}},\ \bibinfo {pages} {372–403} (\bibinfo {year} {2017})}\BibitemShut
  {NoStop}%
\bibitem [{\citenamefont {Hamilton}\ \emph {et~al.}(2016)\citenamefont
  {Hamilton}, \citenamefont {Leskovec},\ and\ \citenamefont
  {Jurafsky}}]{hamilton2016diachronic}%
  \BibitemOpen
  \bibfield  {author} {\bibinfo {author} {\bibfnamefont {W.~L.}\ \bibnamefont
  {Hamilton}}, \bibinfo {author} {\bibfnamefont {J.}~\bibnamefont {Leskovec}},\
  and\ \bibinfo {author} {\bibfnamefont {D.}~\bibnamefont {Jurafsky}},\
  }\bibfield  {title} {\bibinfo {title} {Diachronic word embeddings reveal
  statistical laws of semantic change},\ }in\ \href
  {https://doi.org/10.18653/v1/P16-1141} {\emph {\bibinfo {booktitle}
  {Proceedings of the 54th Annual Meeting of the Association for Computational
  Linguistics (Volume 1: Long Papers)}}}\ (\bibinfo {year} {2016})\ pp.\
  \bibinfo {pages} {1489--1501}\BibitemShut {NoStop}%
\bibitem [{\citenamefont {Schlechtweg}\ \emph {et~al.}(2020)\citenamefont
  {Schlechtweg}, \citenamefont {McGillivray}, \citenamefont {Hengchen},
  \citenamefont {Dubossarsky},\ and\ \citenamefont
  {Tahmasebi}}]{schlechtweg-etal-2020-semeval}%
  \BibitemOpen
  \bibfield  {author} {\bibinfo {author} {\bibfnamefont {D.}~\bibnamefont
  {Schlechtweg}}, \bibinfo {author} {\bibfnamefont {B.}~\bibnamefont
  {McGillivray}}, \bibinfo {author} {\bibfnamefont {S.}~\bibnamefont
  {Hengchen}}, \bibinfo {author} {\bibfnamefont {H.}~\bibnamefont
  {Dubossarsky}},\ and\ \bibinfo {author} {\bibfnamefont {N.}~\bibnamefont
  {Tahmasebi}},\ }\bibfield  {title} {\bibinfo {title} {{S}em{E}val-2020 task
  1: Unsupervised lexical semantic change detection},\ }in\ \href
  {https://doi.org/10.18653/v1/2020.semeval-1.1} {\emph {\bibinfo {booktitle}
  {Proceedings of the Fourteenth Workshop on Semantic Evaluation}}}\ (\bibinfo
  {publisher} {International Committee for Computational Linguistics},\
  \bibinfo {address} {Barcelona (online)},\ \bibinfo {year} {2020})\ pp.\
  \bibinfo {pages} {1--23}\BibitemShut {NoStop}%
\bibitem [{\citenamefont {Sinha}\ \emph {et~al.}(2015)\citenamefont {Sinha},
  \citenamefont {Shen}, \citenamefont {Song}, \citenamefont {Ma}, \citenamefont
  {Eide}, \citenamefont {Hsu},\ and\ \citenamefont {Wang}}]{sinha2015overview}%
  \BibitemOpen
  \bibfield  {author} {\bibinfo {author} {\bibfnamefont {A.}~\bibnamefont
  {Sinha}}, \bibinfo {author} {\bibfnamefont {Z.}~\bibnamefont {Shen}},
  \bibinfo {author} {\bibfnamefont {Y.}~\bibnamefont {Song}}, \bibinfo {author}
  {\bibfnamefont {H.}~\bibnamefont {Ma}}, \bibinfo {author} {\bibfnamefont
  {D.}~\bibnamefont {Eide}}, \bibinfo {author} {\bibfnamefont {B.-j.~P.}\
  \bibnamefont {Hsu}},\ and\ \bibinfo {author} {\bibfnamefont {K.}~\bibnamefont
  {Wang}},\ }\bibfield  {title} {\bibinfo {title} {An overview of microsoft
  academic service (mas) and applications},\ }in\ \href
  {https://doi.org/10.1145/2740908.2742839} {\emph {\bibinfo {booktitle}
  {Proceedings of the 24th International Conference on World Wide Web}}}\
  (\bibinfo {year} {2015})\ p.\ \bibinfo {pages} {243–246}\BibitemShut
  {NoStop}%
\bibitem [{\citenamefont {Mikolov}\ \emph
  {et~al.}(2013{\natexlab{b}})\citenamefont {Mikolov}, \citenamefont
  {Sutskever}, \citenamefont {Chen}, \citenamefont {Corrado},\ and\
  \citenamefont {Dean}}]{mikolov2013distributed}%
  \BibitemOpen
  \bibfield  {author} {\bibinfo {author} {\bibfnamefont {T.}~\bibnamefont
  {Mikolov}}, \bibinfo {author} {\bibfnamefont {I.}~\bibnamefont {Sutskever}},
  \bibinfo {author} {\bibfnamefont {K.}~\bibnamefont {Chen}}, \bibinfo {author}
  {\bibfnamefont {G.~S.}\ \bibnamefont {Corrado}},\ and\ \bibinfo {author}
  {\bibfnamefont {J.}~\bibnamefont {Dean}},\ }\bibfield  {title} {\bibinfo
  {title} {Distributed representations of words and phrases and their
  compositionality},\ }in\ \href@noop {} {\emph {\bibinfo {booktitle} {NIPS}}}\
  (\bibinfo {year} {2013})\BibitemShut {NoStop}%
\bibitem [{\citenamefont {{\v R}eh{\r u}{\v r}ek}\ and\ \citenamefont
  {Sojka}(2010)}]{rehurek_lrec}%
  \BibitemOpen
  \bibfield  {author} {\bibinfo {author} {\bibfnamefont {R.}~\bibnamefont {{\v
  R}eh{\r u}{\v r}ek}}\ and\ \bibinfo {author} {\bibfnamefont {P.}~\bibnamefont
  {Sojka}},\ }\bibfield  {title} {\bibinfo {title} {Software framework for
  topic modelling with large corpora},\ }in\ \href@noop {} {\emph {\bibinfo
  {booktitle} {Proceedings of the LREC 2010 Workshop on New Challenges for NLP
  Frameworks}}}\ (\bibinfo {year} {2010})\ pp.\ \bibinfo {pages}
  {45--50}\BibitemShut {NoStop}%
\bibitem [{\citenamefont {Lyu}\ and\ \citenamefont {Ke}(2025)}]{coderepo}%
  \BibitemOpen
  \bibfield  {author} {\bibinfo {author} {\bibfnamefont {Z.}~\bibnamefont
  {Lyu}}\ and\ \bibinfo {author} {\bibfnamefont {Q.}~\bibnamefont {Ke}},\
  }\href@noop {} {\bibinfo {title} {Diachronic periodical embeddings reveal the
  evolution of science. {GitHub}.
  https://github.com/netknowledge/diachronic-p2v}} (\bibinfo {year} {2025}),\
  \bibinfo {note} {deposited 28 April 2025}\BibitemShut {NoStop}%
\bibitem [{\citenamefont {Sch{\"o}nemann}(1966)}]{schonemann1966procrustes}%
  \BibitemOpen
  \bibfield  {author} {\bibinfo {author} {\bibfnamefont {P.~H.}\ \bibnamefont
  {Sch{\"o}nemann}},\ }\bibfield  {title} {\bibinfo {title} {A generalized
  solution of the orthogonal procrustes problem},\ }\href@noop {} {\bibfield
  {journal} {\bibinfo  {journal} {Psychometrika}\ }\textbf {\bibinfo {volume}
  {31}},\ \bibinfo {pages} {1} (\bibinfo {year} {1966})}\BibitemShut {NoStop}%
\bibitem [{\citenamefont {Maaten}\ and\ \citenamefont
  {Hinton}(2008)}]{maaten2008visualizing}%
  \BibitemOpen
  \bibfield  {author} {\bibinfo {author} {\bibfnamefont {L.~v.~d.}\
  \bibnamefont {Maaten}}\ and\ \bibinfo {author} {\bibfnamefont
  {G.}~\bibnamefont {Hinton}},\ }\bibfield  {title} {\bibinfo {title}
  {Visualizing data using t-sne},\ }\href@noop {} {\bibfield  {journal}
  {\bibinfo  {journal} {Journal of Machine Learning Research}\ }\textbf
  {\bibinfo {volume} {9}},\ \bibinfo {pages} {2579} (\bibinfo {year}
  {2008})}\BibitemShut {NoStop}%
\end{thebibliography}
\end{document}